\documentclass{LMCS}

\usepackage{hyperref}
\usepackage{enumerate}

\usepackage{ifpdf}
\ifpdf
\usepackage[matrix,arrow,tips,curve]{xy}
\else
\usepackage[dvips,matrix,arrow,tips,curve]{xy}
\fi

\usepackage{xspace}
\usepackage{stmaryrd}
\usepackage{amssymb}

\theoremstyle{definition}
\newtheorem{notation}[thm]{Notation}
\newtheorem{example}[thm]{Example}

\SelectTips{cm}{10}

\usepackage[english]{babel}

\newcommand{\trsp}[3]{\mathcal{#1} = (#2, #3)}

\newcommand{\rew}{\rightarrow}
\newcommand{\rewt}{\rightarrow^*}
\newcommand{\trewt}{\twoheadrightarrow}
\newcommand{\trewtp}[1]{\twoheadrightarrow^{#1}}

\newcommand{\emmy}{\mathord{\sslash}}

\newcommand{\pos}[1]{\mathcal{P}os(#1)}

\newcommand{\rs}[1]{root(#1)}

\newcommand{\natnum}{\mathbb{N}}

\newcommand{\seper}{\; | \;}

\newcommand{\redpos}[1]{p_{#1}}

\newcommand{\apath}{\Pi}

\newcommand{\ulam}{\underline{\lambda}}

\newcommand{\iLC}{i$\lambda$c\xspace}

\newcommand{\dev}{\Rightarrow}

\newcommand{\pmap}{\varepsilon}
\newcommand{\pmaph}[1]{\pmap_{#1}}
\newcommand{\pmapp}[2]{\pmap_{#1}({#2})}

\newcommand{\pme}{\mu}
\newcommand{\pmep}[2]{\pme_{#1}({#2})}

\newcommand{\fscd}[3]{{#1}_0 \dev^{\mathcal{#2}_1} {#1}_1 \dev^{\mathcal{#2}_2} \cdots \dev^{\mathcal{#2}_{#3}} {#1}_{#3}}
\newcommand{\fscdp}[3]{{#1}'_0 \dev^{\mathcal{#2}'_1} {#1}'_1 \dev^{\mathcal{#2}'_2} \cdots \dev^{\mathcal{#2}'_{#3}} {#1}'_{#3}}

\newcommand{\Parg}[1]{\mathcal{P}_{\mathrm{#1}}\xspace}
\newcommand{\Ptrue}[0]{\Parg{true}}
\newcommand{\Pneeded}[0]{\Parg{needed}}
\newcommand{\Poutermost}[0]{\Parg{outermost}}

\def\doi{6 (1:7) 2010}
\lmcsheading%
{\doi}
{1--35}
{}
{}
{Oct.~28, 2008}
{Feb.~26, 2010}
{}   

\begin{document}

\title[Infinitary Combinatory Reduction Systems: Normalising Reduction Strategies]{Infinitary Combinatory Reduction Systems: \\
Normalising Reduction Strategies\rsuper*}

\author[J.~Ketema]{Jeroen Ketema\rsuper a}
\address{{\lsuper a}Research Institute of Electrical Communication, Tohoku University\\
2-1-1 Katahira, Aoba-ku, Sendai 980-8577, Japan}
\email{jketema@nue.riec.tohoku.ac.jp}
\thanks{{\lsuper a}This author was partially funded by the Netherlands Organisation for Scientific Research (NWO) under FOCUS/BRICKS grant number 642.000.502.}

\author[J.~G.~Simonsen]{Jakob Grue Simonsen\rsuper b}
\address{{\lsuper b}Department of Computer Science, University of Copenhagen (DIKU)\\
Universitetsparken 1, 2100 Copenhagen \O, Denmark}
\email{simonsen@diku.dk}

\keywords{term rewriting, higher-order computation, combinatory reduction systems, lambda-calculus, infinite computation, reduction strategies, normal forms}
\subjclass{D.3.1, F.3.2, F.4.1, F.4.2}
\titlecomment{{\lsuper*}Parts of this paper have previously appeared as \cite{JJ05b}}

\begin{abstract}
We study normalising reduction strategies for infinitary Combinatory Reduction Systems (iCRSs). We prove that all fair, outermost-fair, and need\-ed-fair strategies are normalising for orthogonal, fully-extended iCRSs. These facts properly generalise a number of results on normalising strategies in first-order infinitary rewriting and provide the first examples of normalising strategies for infinitary $\lambda$-calculus.

\end{abstract}

\maketitle

\tableofcontents

\section{Introduction}

This paper is part of a series outlining the theory and basic results of \emph{infinitary Combinatory Reduction Systems (iCRSs)}, the first notion of infinitary \emph{higher-order} term rewriting. Preliminary papers \cite{JJ05a,JJ05b} have established basic notions of terms and complete developments. The present paper is devoted to the study of reductions to normal form, in particular the study of normalising reduction strategies.

The purpose is to extend infinitary term rewriting to encompass higher-order rewrite systems.
This allows us, for instance, to reason about the behaviour of the well-known $\mathtt{map}$ functional 
when it is applied to infinite lists. The $\mathtt{map}$ functional
and the usual constructors and destructors for lists can be represented by the below iCRS:
\begin{align*}
\mathtt{map}([z]F(z),\mathtt{cons}(X,XS)) & \rew
\mathtt{cons}(F(X),\mathtt{map}([z]F(z),XS)) \\
\mathtt{map}([z]F(z),\mathtt{nil}) & \rew \mathtt{nil}\\
\mathtt{hd}(\mathtt{cons}(X,XS)) & \rew X \\
\mathtt{tl}(\mathtt{cons}(X,XS)) & \rew XS
\end{align*}

Evaluation in systems as the above usually follows a certain \emph{reduction strategy}. Intuitively, a reduction strategy is an algorithm that, for a given term, chooses the redex to contract if several redexes appear in the term. A reduction strategy is \emph{normalising} if following the strategy will, if at all possible, eventually lead to a normal form: a term containing no redexes.
Normalising strategies have previously been studied in the context of 
ordinary (finitary) term rewriting \cite{HL91,T03_OV}
and in the context of first-order infinitary rewriting \cite{KKSV95}. 

In general, the methods for proving normalisation of a number of strategies from finitary rewriting could be lifted to the first-order infinitary setting without severe difficulty. In the higher-order case, the fact that rewrite rules may nest subterms combined with the desire to treat rewrite rules with right-hand sides that are possibly infinite, conspires to render the methods from first-order infinitary rewriting and, to a great extent, from higher-order finitary rewriting unusable. It turns out, however, that a particular
technique due to van Oostrom \cite{O99} is usable in somewhat modified form
for proving normalisation in the higher-order infinitary setting.

\subsubsection*{Contributions}

The main contributions of this paper are:

\begin{enumerate}[$\bullet$]

\item the result that any fair, outermost-fair, and needed-fair strategy is normalising for orthogonal, fully-extended iCRSs, and

\item the development of novel techniques for treating sequences of complete developments, especially the methods of \emph{essential rewrite steps} and \emph{emaciated projections}.

\end{enumerate}

The first contribution properly generalises identical results on normalising strategies known from first-order infinitary rewriting \cite{KKSV95}; it does so as infinitary (first-order) term rewriting systems (iTRSs) can be regarded as special cases of iCRSs. Furthermore, the first contribution also provides the first normalising strategies for infinitary $\lambda$-calculus (\iLC) \cite{KKSV97}, as \iLC can be viewed as a specific example of an iCRS.

The second contribution, apart from its added value per se, facilitates proofs of confluence properties in orthogonal, fully-extended iCRSs that appear in a companion paper on confluence \cite{paper_iv}.

Note that we do not prove the external-fair, parallel-outermost and depth-increasing strategies known from first-order infinitary rewriting \cite{KKSV95} to be normalising. However, in related research, the first author does show that any needed strategy is normalising \cite{K08}; the author's proof builds on techniques developed in the present paper.

\subsubsection*{Layout of the paper}

Section~\ref{sec:preliminaries} recapitulates basic definitions, Section~\ref{sec:overview} provides an overview of the proof techniques and a roadmap to the results, Section~\ref{sec:ess} introduces the concept of an \emph{essential redex}, the technical fulcrum of the paper, Section~\ref{sec:normal} proves the main results concerning normalising strategies, and Section~\ref{sec:conclusion} concludes and provides pointers for further work.

\section{Preliminaries}
\label{sec:preliminaries}

We presuppose a working knowledge of the basics of ordinary finitary term rewriting \cite{T03}. The basic theory of infinitary Combinatory Reduction Systems
has been laid out in \cite{JJ05a,JJ05b}, and we give only the briefest of definitions in this section. Full proofs of all results may be
found in the above-mentioned papers. Moreover, the reader familiar with \cite{paper_i} may safely skip this section as it is essentially an abstract of that paper.

Throughout, infinitary Term Rewriting Systems are invariably abbreviated as iTRSs and infinitary $\lambda$-calculus is abbreviated as \iLC. Moreover, we denote the first infinite ordinal by $\omega$, and arbitrary ordinals by $\alpha$, $\beta$, $\gamma$, and so on. We use $\natnum$ to denote the set of natural numbers, starting from zero.

\subsection{Terms, meta-terms, and positions}

We assume a signature $\Sigma$, each element of which has finite arity. We also assume a countably infinite set of variables and, for each finite arity, a countably infinite set of meta-variables of that arity. Countably infinite sets suffice, given that we can employ `Hilbert hotel'-style renaming. 

The (infinite) \emph{meta-terms} are defined informally in a top-down fashion by the following rules, where $s$ and $s_1$, \ldots, $s_n$ are again meta-terms:
\begin{enumerate}[(1)]
\item
each variable $x$ is a meta-term,
\item
if $x$ is a variable and $s$ is a finite meta-term, then $[x] s$ is a meta-term,
\item
if $Z$ is a meta-variable of arity $n$, then $Z(s_1, \ldots, s_n)$ is a meta-term,
\item
if $f \in \Sigma$ has arity $n$, then $f(s_1, \ldots, s_n)$ is a meta-term.
\end{enumerate}

\noindent We consider meta-terms modulo $\alpha$-equivalence.

A meta-term of the form $[x] s$ is called an
\emph{abstraction}. Each occurrence of the variable $x$ in $s$ is
\emph{bound} in $[x]s$, and each subterm of $s$ is said to occur in the
\emph{scope} of the abstraction. If $s$ is a meta-term, we denote by $\rs{s}$ the root symbol of $s$. Following the definition of meta-terms, we define $\rs{x} = x$, $\rs{[x]s} = [x]$, $\rs{Z(s_1, \ldots, s_n)} = Z$, and $\rs{f(s_1, \ldots, s_n)} = f$.

The set of \emph{terms} is defined as the set of all meta-terms without meta-variables. Moreover, a \emph{context} is defined as a meta-term over $\Sigma \cup \{ \Box \}$ where $\Box$ is a fresh nullary function symbol and a \emph{one-hole} context is a context in which precisely one $\Box$ occurs. If $C[\Box]$ is a one-hole context and $s$ is a term, we obtain a term by replacing $\Box$ by s; the new term is denoted by $C[s]$.

Replacing a hole in a context does not avoid the capture of free variables: A free variable $x$ in $s$ is bound by an abstraction over $x$ in $C[\Box]$ in case $\Box$ occurs in the scope of the abstraction. This behaviour is not obtained automatically when working modulo $\alpha$-equivalence: It is \emph{always} possible find a representative from the $\alpha$-equivalence class of $C[\Box]$ that does not capture the free variables in $s$. Therefore, we will always work with \emph{fixed} representatives from $\alpha$-equivalence classes of contexts. This convention ensures that variables will be captured properly.

\begin{rem}
Capture avoidance is disallowed for contexts as we do not want to lose variable bindings over rewrite steps in case: (i) an abstraction occurs in a context, and (ii) a variable bound by the abstraction occurs in a subterm being rewritten. Note that this means that the representative employed as the context must already be fixed \emph{before} performing the actual rewrite step.

As motivation, consider $\lambda$-calculus: In the term
$\lambda x . (\lambda y . x) z$, contracting the redex 
inside the context $\lambda x . \Box$ yields
$\lambda x.x$, whence the substitution rules for contexts should
be such that 
\[
(\lambda x. \Box)\{(\lambda y . x) z / \Box\}
\rew_\beta \lambda x.x \, .
\]
 If capture avoidance
were in effect for contexts, we would have an $\alpha$-conversion
in the rewrite step, whence
\[
(\lambda x .\Box)\{(\lambda y . x) z / \Box\}
\rew_\beta \lambda w . x \, ,
\]
which is clearly wrong.
\end{rem}

Formally, meta-terms are defined by taking the metric completion of the set of finite meta-terms, the set inductively defined by the rules presented earlier. The distance between two terms is either taken as $0$, if the terms are $\alpha$-equivalent, or as $2^{-k}$ with $k$ the minimal depth at which the terms differ, also taking into account $\alpha$-equivalence. By definition of metric completion, the set of finite meta-terms is a subset of the set of meta-terms. Moreover, the metric on finite meta-terms extends uniquely to a metric on meta-terms.

\begin{exa}
Any finite meta-term, e.g.\ $[x]Z(x,f(x))$, is a meta-term. We also have that $Z'(Z'(Z'(\ldots)))$ is a meta-term, as is $Z_1([x_1]x_1,Z_2([x_2]x_2,\ldots))$.

The meta-terms $[x]Z(x,f(x))$ and $[y]Z(y,f(y))$ have distance $0$ and the meta-terms $[x]Z(x,f(x))$ and $[y]Z(y,f(z))$ have distance $\frac{1}{8}$.
\end{exa}

Positions of meta-terms are defined by considering such terms in a top-down fashion. Given a meta-term $s$, its \emph{set of positions}, denoted $\pos{s}$, is the set of finite strings over $\natnum$, with $\epsilon$ the empty string, such that:
\begin{enumerate}[(1)]
\item
if $s = x$ for some variable $x$, then $\pos{s} = \{ \epsilon \}$,
\item
if $s = [x] t$, then $\pos{s} = \{ \epsilon \} \cup \{ 0 \cdot p \seper p \in \pos{t} \}$,
\item
if $s = Z(t_1, \ldots, t_n)$, then $\pos{s} = \{ \epsilon \} \cup \{ i \cdot p \seper 1 \leq i \leq n, \, p \in \pos{t_i} \}$,
\item
if $s = f(t_1, \ldots, t_n)$, then $\pos{s} = \{ \epsilon \} \cup \{ i \cdot p \seper 1 \leq i \leq n, \, p \in \pos{t_i} \}$.
\end{enumerate}

The \emph{depth} of a position $p$, denoted $\vert p \vert$, is the
number of characters in $p$. Given $p, \, q \in \pos{s}$, we write $p \leq q$
and say that $p$ is a \emph{prefix} of $q$, if there exists an $r \in \pos{s}$ such that $p \cdot r = q$. If
$r \not = \epsilon$, we also write $p < q$ and say that the prefix is
\emph{strict}. Moreover, if neither $p \leq q$ nor $q \leq p$, we say that $p$
and $q$ are \emph{parallel}, which we write as $p \parallel q$. 

We denote by $s|_p$ the subterm of $s$ that \emph{occurs at position} $p \in \pos{s}$. Moreover, if $q \in \pos{s}$ and $p < q$, we say that the subterm at position $p$ occurs \emph{above} $q$. Finally, if $p > q$, then we say that the subterm occurs \emph{below} $q$.

Below we introduce a restriction on meta-terms called the \emph{finite chains property}, which enforces the proper behaviour of valuations. Intuitively, a \emph{chain} is a sequence of contexts in a meta-term occurring `nested right below each other'.
\begin{defi}
Let $s$ be a meta-term.  A \emph{chain} in $s$ is a sequence
of (context, position)-pairs $(C_i[\Box],p_i)_{i < \alpha}$, with
$\alpha \leq \omega$, such that for each $(C_i[\Box], p_i)$:
\begin{enumerate}[(1)]
\item
if $i + 1 < \alpha$, then $C_i[\Box]$ has \emph{one} hole and $C_i[t_i] = s|_{p_i}$ for some term $t_i$, and
\item
if $i + 1 = \alpha$, then $C_i[\Box]$ has \emph{no} holes and $C_i[\Box] = s|_{p_i}$,
\end{enumerate}
and such that $p_{i+1} = p_i \cdot q_i$ for all $i + 1 < \alpha$ where $q_i$ is the position of the hole in $C_i[\Box]$.

If $\alpha < \omega$, respectively
$\alpha = \omega$, then the chain is called \emph{finite},
respectively \emph{infinite}.
\end{defi}
\noindent Observe that at most one $\Box$ occurs in any context $C_i[\Box]$ in a chain. In fact, $\Box$ only occurs in $C_i[\Box]$ if $i + 1 < \alpha$; if $i + 1 = \alpha$, we have $C_i[\Box] = s|_{p_i}$.

\subsection{Valuations}

We next define valuations, the iCRS analogue of substitutions as defined for iTRSs and \iLC. As it turns out, the most straightforward and liberal definition of meta-terms has rather poor properties: Applying a valuation
need not necessarily yield a well-defined term. Therefore, we also introduce an important restriction on meta-terms: the \emph{finite chains property}. This property will also prove crucial in obtaining positive results later in the paper.

Essentially, the definitions are the same as in the case of CRSs \cite{KOR93,T03_R}, except that the interpretation of the definition is top-down (due to the presence of infinite terms and meta-terms). Below, we use $\vec{x}$ and $\vec{t}$ as short-hands for, respectively, the sequences $x_1, \ldots, x_n$ and $t_1, \ldots, t_n$ with $n \geq 0$. Moreover, we assume $n$ fixed in the next two definitions.
\begin{defi}
\label{defsubstitution}
A \emph{substitution} of the terms 
$\vec{t}$ for distinct variables 
$\vec{x}$ in a term $s$, denoted $s[\vec{x} := \vec{t}]$, is defined as:
\begin{enumerate}[(1)]
\item
$x_i[\vec{x} := \vec{t}] = t_i$,
\item
$y[\vec{x} := \vec{t}] = y$, if $y$ does not occur in $\vec{x}$,
\item
$([y]s')[\vec{x} := \vec{t}] = [y](s'[\vec{x} := \vec{t}])$,
\item
$f(s_1, \ldots, s_m)[\vec{x} := \vec{t}] = f(s_1[\vec{x} := \vec{t}],
\ldots, s_m[\vec{x} := \vec{t}])$.
\end{enumerate}
The above definition implicitly takes into account the usual variable
convention \cite{B85} in the third clause to avoid the binding of free
variables by the abstraction. We now define substitutes (adopting this
name from Kahrs \cite{K93}) and valuations.
\end{defi}

\begin{defi}
\label{defsubstitute}
An \emph{$n$-ary substitute} is a mapping denoted $\ulam x_1, \ldots, x_n . s$ or $\ulam \vec{x} . s$, with $s$ a term, such that:
\begin{equation}
\label{pbetaeq}
(\ulam \vec{x} . s)(t_1, \ldots, t_n) = s[\vec{x} := \vec{t}] \, .
\end{equation}
\end{defi}

The intention of a substitute is to ensure that proper `housekeeping' 
of substitutions is observed when performing a rewrite step. Reading Equation (\ref{pbetaeq}) from left to right yields a rewrite rule:
\[
(\ulam \vec{x}. s)(t_1, \ldots, t_n) \rew s[\vec{x} := \vec{t}] \, .
\]
The rule can be seen as a \emph{parallel $\beta$-rule}. That is, a variant of the $\beta$-rule from (infinitary) $\lambda$-calculus which simultaneously substitutes multiple variables.

\begin{defi}
\label{defvaluation}
Let $\sigma$ be a function that maps meta-variables to substitutes
such that, for all $n \in \mathbb{N}$, if $Z$ has arity $n$, then so
does $\sigma(Z)$.

A \emph{valuation} induced by $\sigma$ is a relation $\bar{\sigma}$ that takes meta-terms to terms such that:
\begin{enumerate}[(1)]
\item
$\bar{\sigma}(x) = x$,
\item
$\bar{\sigma}([x]s) = [x](\bar{\sigma}(s))$,
\item
$\bar{\sigma}(Z(s_1, \ldots, s_m)) = \sigma(Z)(\bar{\sigma}(s_1), \ldots, \bar{\sigma}(s_m))$,
\item
$\bar{\sigma}(f(s_1, \ldots, s_m)) = f(\bar{\sigma}(s_1), \ldots, \bar{\sigma}(s_m))$.
\end{enumerate}

\end{defi}

Similar to Definition \ref{defsubstitution}, the above definition implicitly takes the variable convention into account, this time in the second clause, to avoid the binding of free variables by the abstraction.

The definition of a valuation yields a straightforward two-step way of applying it to a meta-term: In the first step each subterm of the form $Z(t_1, \ldots, t_n)$ is replaced by a subterm of the form $(\ulam \vec{x}. s)(t_1, \ldots, t_n)$. In the second step Equation \eqref{pbetaeq} is applied to each of these subterms.

In the case of (finite) CRSs, valuations are always (everywhere defined) \emph{maps} taking each meta-term to a unique term \cite[Remark II.1.10.1]{K80}. This is no longer the case when infinite meta-terms are considered. For example, given the meta-term $Z(Z(\ldots Z(\ldots)))$ and applying any map that satisfies $Z \mapsto \ulam x.x$, we obtain $(\ulam x. x)((\ulam x. x)(\ldots (\ulam x. x)(\ldots)))$. Viewing Equation \eqref{pbetaeq} as a rewrite rule, this `$\ulam$-term' reduces only to itself and never to a \emph{term}, as required by the definition of valuations (for more details, see \cite{JJ05a}). To mitigate this problem a subset of the set of meta-terms is introduced in \cite{JJ05a}.
\begin{defi}
Let $s$ be a meta-term.  A \emph{chain of meta-variables} in $s$ is a chain in $s$, written $(C_i[\Box],p_i)_{i < \alpha}$ with $\alpha \leq \omega$, such that for each $i < \alpha$ it is the case that $C_i[\Box] = Z(t_1, \ldots, t_n)$ with $t_j = \Box$ for exactly one $1 \leq j \leq n$.

The meta-term $s$ is said to satisfy the \emph{finite chains property} if no infinite chain of meta-variables occurs in $s$.
\end{defi}

\begin{exa}
The meta-term $[x_1]Z_1([x_2]Z_2(\ldots [x_n]Z_n(\ldots)))$ satisfies the finite chains property. The meta-terms $Z(Z(\ldots Z(\ldots)))$ and $Z_1(Z_2(\ldots Z_n(\ldots)))$ do not.
\end{exa}

From \cite{JJ05a} we now have the following result:
\begin{prop}
\label{prop:metasane}
Let $s$ be a meta-term satisfying the finite chains property and
let $\bar{\sigma}$ a valuation. There is a unique term that is the
result of applying $\bar{\sigma}$ to $s$. \qed
\end{prop}

\subsection{Rewrite rules and reductions}
\label{sec:rules}

Having defined terms and valuations, we move on to define rewrite rules and reductions.

\subsubsection{Rewrite rules}

We give a number of definitions that are direct extensions of the
corresponding definitions from CRS theory.

\begin{defi}
A finite meta-term is a \emph{pattern} if each of its meta-variables has distinct bound variables as its arguments. Moreover, a meta-term is \emph{closed} if all its variables occur bound.
\end{defi}

We next define rewrite rules and iCRSs. The definitions are identical to the definitions in the finite case, with the exception of the restrictions on the right-hand sides of the rewrite rules: The finiteness restriction is lifted and the finite chains property is put in place. 
\begin{defi}
A \emph{rewrite rule} is a pair $(l, r)$, denoted $l \rew r$, where $l$ is a finite meta-term and $r$ is a meta-term, such that:
\begin{enumerate}[(1)]
\item
$l$ is a pattern with a function symbol at the root,
\item
all meta-variables that occur in $r$ also occur in $l$, 
\item
$l$ and $r$ are closed, and
\item
$r$ satisfies the finite chains property.
\end{enumerate}
The meta-terms $l$ and $r$ are called, respectively, the \emph{left-hand side}
and the \emph{right-hand side} of the rewrite rule.

An \emph{infinitary Combinatory Reduction System (iCRS)} is a pair
$\trsp{C}{\Sigma}{R}$ with $\Sigma$ a signature and $R$ a set of rewrite rules.
\end{defi}

With respect to the left-hand sides of rewrite rules, it is always the case that only finite chains of meta-variables occur, as the left-hand sides are finite.

We now define rewrite steps.
\begin{defi}
A \emph{rewrite step} is a pair of terms $(s, t)$, denoted $s \rew t$, adorned with a one-hole context $C[\Box]$, a rewrite rule $l \rew r$, and a valuation $\bar{\sigma}$ such that $s = C[\bar{\sigma}(l)]$ and $t = C[\bar{\sigma}(r)]$.
The term
$\bar{\sigma}(l)$ is called an \emph{$l \rew r$-redex}, or
simply a \emph{redex}. The redex \emph{occurs} at position $p$ and
depth $|p|$ in $s$, where $p$ is the position of the hole in $C[\Box]$.

A position $q$ of $s$ is said to occur in the \emph{redex pattern} of
the redex at position $p$ if $q \geq p$ and if there does
not exist a position $q'$ with $q \geq p \cdot q'$ such that $q'$ is the
position of a meta-variable in $l$.
\end{defi}

For example, $f([x]Z(x),Z') \rightarrow Z(Z')$ is a rewrite rule,
and $f([x]h(x),a)$ rewrites to $h(a)$ by contracting the redex of the rule 
$f([x]Z(x),Z') \rightarrow Z(Z')$ occurring at position $\epsilon$, i.e.\ at the root.

We now mention some standard restrictions on rewrite rules 
that we need later in the paper: 

\begin{defi}
A rewrite rule is \emph{left-linear}, if each meta-variable occurs at
most once in its left-hand side. Moreover, an iCRS is \emph{left-linear}
if all its rewrite rules are.
\end{defi}

\begin{defi}
Let $s$ and $t$ be finite meta-terms that have no meta-variables in common. The meta-term $s$ \emph{overlaps} $t$ if there exists a non-meta-variable position $p \in \pos{s}$ and a valuation $\bar{\sigma}$ such that $\bar{\sigma}(s|_p) = \bar{\sigma}(t)$.

Two rewrite rules \emph{overlap} if their left-hand sides overlap and if the overlap does not occur at the root when two copies of the same rule are considered. An iCRS is \emph{orthogonal} if all its rewrite rules are left-linear and no two (possibly the same) rewrite rules overlap.
\end{defi}

In case the rewrite rules $l_1 \rew r_1$ and $l_2 \rew r_2$ overlap
at position $p$, it follows that $p$ cannot be the position of a
bound variable in $l_1$. If it were, we would obtain for some valuation
$\bar{\sigma}$ and variable $x$ that $\bar{\sigma}(l_1|_p) = x = 
\bar{\sigma}(l_2)$, which would imply that $l_2$ does not have
a function symbol at the root, as required by the definition of
rewrite rules.

Moreover, it is easily seen that if two left-linear rules overlap in an infinite term, there is also a finite term in which they overlap. As left-hand sides are
\emph{finite} meta-terms, we may appeal to standard ways of deeming
CRSs orthogonal by inspection of their rules.
We shall do so informally on several occasions in the
remainder of the paper.

\begin{defi}
\label{def:fully_extended}
A pattern is \emph{fully-extended} \cite{HP96,O96}, if, for each of
its meta-variables $Z$ and each abstraction $[x]s$ having an
occurrence of $Z$ in its scope, $x$ is an argument of that occurrence
of $Z$. Moreover, a rewrite rule is \emph{fully-extended} if its
left-hand side is and an iCRS is \emph{fully-extended} if all its
rewrite rules are.
\end{defi}

\begin{exa}
The pattern $f(g([x]Z(x)))$ is fully-extended. Hence, so is the rewrite
rule $f(g([x]Z(x))) \rightarrow h([x]Z(x))$. The pattern $g([x]f(Z(x),Z'))$, with $Z'$ occurring in the scope of the abstraction $[x]$, is not fully-extended as $x$ does not occur as an argument of $Z'$.
\end{exa}

\subsubsection{Transfinite reductions}

We can now define transfinite reductions.
The definition is equivalent to those for iTRSs and \iLC \cite{KKSV95,KKSV97}.
\begin{defi}
A \emph{transfinite reduction} with domain $\alpha > 0$ is a sequence
of terms $(s_\beta)_{\beta < \alpha}$ adorned with a rewrite step
$s_\beta \rew s_{\beta + 1}$ for each $\beta + 1 < \alpha$. In case $\alpha = \alpha' + 1$,
the reduction is \emph{closed} and of length $\alpha'$. In case
$\alpha$ is a limit ordinal, the reduction is called \emph{open} and
of length $\alpha$. The reduction is \emph{weakly continuous} or
\emph{Cauchy continuous} if, for every limit ordinal $\gamma < \alpha$, the distance between $s_\beta$ and $s_\gamma$ tends to $0$ as $\beta$ approaches $\gamma$ from below. The reduction is \emph{weakly convergent} or \emph{Cauchy convergent} if it is weakly continuous and closed.
\end{defi}
\noindent Intuitively, an open transfinite reduction is lacking a well-defined final term, while a closed reduction does have such a term. 

As in \cite{KKSV95,KKSV97,T03_KV}, we prefer to reason about strongly
convergent reductions.

\begin{defi}
Let $(s_\beta)_{\beta < \alpha}$ be a transfinite reduction. For each
rewrite step $s_\beta \rew s_{\beta + 1}$, let $d_\beta$ denote the
depth of the contracted redex. The reduction is \emph{strongly
  continuous} if it is weakly continuous and if, for every limit ordinal $\gamma < \alpha$, the depth $d_\beta$ tends to infinity as $\beta$ approaches $\gamma$ from below. The reduction is \emph{strongly convergent} if strongly continuous and closed.
\end{defi}

\begin{exa}
\label{ex:wconv}
Consider the rewrite rule $f([x]Z(x)) \rew Z(f([x]Z(x)))$ and observe that $f([x]x) \rew f([x]x)$. Define $s_\beta = f([x]x)$ for all $\beta < \omega \cdot 2$. The reduction $(s_\beta)_{\beta < \omega \cdot 2}$, where in each step we contract the redex at the root, is open and weakly continuous. Adding the term $f([x]x)$ to the end of the reduction yields a weakly convergent reduction. Both reductions are of length $\omega \cdot 2$.

The above reduction is not strongly continuous as all contracted redexes occur at the root, i.e.\ at depth $0$. 
In addition, it cannot be extended to a strongly convergent reduction.
However, the reduction
\[
f([x]g(x)) \rew g(f([x]g(x)) \rew \cdots \rew g^n(f([x]g(x))) \rew g^{n + 1}(f([x]g(x))) \rew \cdots
\]
is open and strongly continuous. Extending the reduction with the term $g^\omega$, where $g^\omega$ is shorthand for the infinite term $g(g(\ldots g(\ldots)))$, yields a strongly convergent reduction. Both reductions are of length $\omega$.
\end{exa}

\begin{notation}
By $s \trewtp{\alpha} t$, respectively $s \trewtp{\leq \alpha} t$, we
denote a \emph{strongly convergent} reduction of ordinal
length $\alpha$, respectively of ordinal length at most
$\alpha$. By $s \trewt t$ we denote a \emph{strongly convergent}
reduction of arbitrary ordinal length and by $s \rewt t$
we denote a reduction of finite length.
Reductions are usually ranged over by
capital letters such as $D$, $E$, $S$, and $T$.
The concatenation of reductions $S$ and $T$ 
is denoted by $S ; T$.
\end{notation}

Note that the concatenation of any finite number of strongly convergent reductions yields a strongly convergent reduction. For strongly convergent reductions, the following is proved in \cite{JJ05a}.
\begin{lem}
\label{depthlem}
If $s \trewt t$, then the number of steps contracting redexes at depths less than $d \in \natnum$ is finite for any $d$ and $s \trewt t$ has countable length.
\end{lem}

The following result \cite{JJ05a} shows that, as in other forms of infinitary rewriting, reductions can always be `compressed' to have length at most $\omega$:

\begin{thm}[Compression]
\label{the:compression}
For every fully-extended, left-linear iCRS, if $s \trewtp{\alpha} t$, then $s \trewtp{\leq \omega} t$. \qed
\end{thm}

\subsubsection{Descendants and residuals}
\label{sec:desc}

The twin notions of \emph{descendants} and \emph{residuals} formalise, respectively, ``what happens'' to positions and redexes across reductions. Across a rewrite step, the only positions that can have descendants are those that occur outside the redex pattern of the contracted redex and that are not positions of the variables bound by abstractions in the redex pattern. Across a reduction, the definition of descendants follows the notion of a descendant across a rewrite step, employing strong convergence in the limit ordinal case. We do not appeal to further details of the definitions in the remainder of this paper and these details are hence omitted. For the full definitions we refer the reader to \cite{JJ05a}.

\begin{notation}
Let $s \trewt t$. Assume $P \subseteq \pos{s}$ and $\mathcal{U}$ a set of redexes in $s$. We denote the descendants of $P$ across $s \trewt t$ by $P/(s \trewt t)$ and the residuals of $\mathcal{U}$ across $s \trewt t$ by $\mathcal{U}/(s \trewt t)$. Moreover, if $P = \{ p \}$ and $\mathcal{U} = \{ u \}$, then we also write $p/(s \trewt t)$ and $u/(s \trewt t)$. Finally, if $s \trewt t$ consists of a single step contracting a redex $u$, then we sometimes write~$\mathcal{U}/u$.
\end{notation}

\subsection{Developments}

We need some basic facts about developments which we recapitulate~now.

Assuming in the remainder of this section that every iCRS is \emph{orthogonal} and that $s$ is a term and $\mathcal{U}$ a set of redexes in $s$, we first define developments:
\begin{defi}
A \emph{development} of $\mathcal{U}$ is a strongly convergent reduction
such that each step contracts a residual of a redex in $\mathcal{U}$. A
development $s  \trewt t$ is called \emph{complete} if $\mathcal{U}/(s \trewt
t) = \emptyset$. Moreover, a development is called $\emph{finite}$ if
$s \trewt t$ is finite.
\end{defi}

A complete development of a set of redexes does not necessarily exist in the infinite case. Consider for example the rule $f(Z) \rew Z$ and the term $f^\omega$. The set of all redexes in $f^\omega$ does not have a complete development: After any (partial) development a residual of a redex in $f^\omega$ always remains at the root of the resulting term. Hence, any complete development will have an infinite number of root-steps and thus will not be strongly convergent.

Although complete developments do not always exist, the following results can still be obtained \cite{JJ05b}, where we write $s \dev^\mathcal{U} t$ for the reduction $s \trewt t$ if it is a complete development of the set of redexes $\mathcal{U}$ in $s$.

\begin{lem}
\label{cdacrossd}
If $\mathcal{U}$ has a complete development and if $s \trewt t$ is a (not necessarily complete) development of $\mathcal{U}$, then $\mathcal{U}/(s \trewt t)$ has a complete development. \qed
\end{lem}

\begin{lem}
\label{lem:finite_dev}
Let $s$ be a term and $\mathcal{U}$ a set of redexes in $s$. If $\mathcal{U}$ is finite, then it has a finite complete development. \qed
\end{lem}

\begin{prop}
\label{prop:finite_over_normal}
Let $\mathcal{U}$ and $\mathcal{V}$ be sets of redexes in $s$
such that $\mathcal{U}$ has a complete development $s \dev t$ and 
$\mathcal{V}$ is finite. The following diagram commutes:
\[
\xymatrix{
s \ar@{=>}[r]^{\mathcal{V}} \ar@{=>}[d]^{\mathcal{U}} & t'
\ar@{=>}[d]^{\mathcal{U}/(s \dev^\mathcal{V} t')} \\
t \ar@{=>}[r]_{\mathcal{V}/(s \dev^\mathcal{U} t)} & s'
}
\]
\end{prop}

Finally, if the reduction $D$ consists of a finite number of complete developments $\fscd{s}{U}{n}$ and if $s_0 \rew t_0$ contracts a redex $u$, then $D / u$ denotes the reduction $\fscd{t}{V}{n}$, where $\mathcal{V}_i = \mathcal{U}_i / (s_{i - 1} \dev t_{i - 1})$ for all $0 < i \leq n$ with $s_{i - 1} \dev t_{i - 1}$ a complete development of the residuals of $u$ in $s_{i - 1}$. Written as a diagram:
\[
\xymatrix{
s_0 \ar@{=>}[r]^{\mathcal{U}_1} \ar[d]^{u}
  & s_1 \ar@{=>}[r]^{\mathcal{U}_2} \ar@{=>}[d]
  & \cdot \ar@{.}[r] \ar@{=>}[d]
  & \cdot \ar@{=>}[r]^{\mathcal{U}_n}
  & s_n \ar@{=>}[d]^{u/D} \\
t_0 \ar@{=>}[r]_{\mathcal{V}_1}
  & t_1 \ar@{=>}[r]_{\mathcal{V}_2}
  & \cdot \ar@{.}[r]
  & \cdot \ar@{=>}[r]_{\mathcal{V}_n}
  & t_n
}
\]
The existence of $D / u$ depends on the existence of the complete developments that define the reduction. By Proposition \ref{prop:finite_over_normal}, existence is guaranteed in case $\mathcal{U}_i$ is finite for all $0 < i < n$.

\subsection{Paths and finite jumps}
\label{sec:pathsjumps}

To support the technique of essential rewrite steps below, we use the technique of paths and finite jumps. This technique, which we lifted from \cite{T03_KV} in \cite{JJ05a}, can be used to reason about developments in the infinitary case. In particular, the technique yields a necessary and sufficient characterisation of those sets of redexes that admit complete developments (Theorem \ref{fjdt} and Lemma \ref{compimplfinite} below). The proofs of the results in this section can be found in \cite{JJ05a}; proofs of auxiliary results used in \cite{JJ05a} but omitted there may be found in Appendix \ref{app:proof_III}.

Assuming a term $s$ in an \emph{orthogonal} iCRS and a set $\mathcal{U}$ of redexes in $s$, we first define paths and path projections, where we denote by $\redpos{u}$ the position of the redex $u$ in $s$. Moreover, we say that a variable $x$ is \emph{bound by a redex} $u$ if $x$ is bound by an abstraction $[x]$ which occurs in the left-hand side of the rewrite rule employed in $u$.
\begin{defi}
A \emph{path} of $s$ with respect to $\mathcal{U}$ is a sequence of alternating nodes and edges. Each node is labelled either $(s, p)$ with $p \in \pos{s}$ or $(r, p, \redpos{u})$ with $r$ the right-hand side of a rewrite rule, $p \in \pos{r}$, and $u \in \mathcal{U}$. Each edge is directed and either unlabelled or labelled with an element of $\natnum$.

Every path starts with a node labelled $(s, \epsilon)$. If a node $n$ of a path is labelled $(s, p)$ and has an outgoing edge to a node $n'$, then:
\begin{enumerate}[(1)]
\item
if $s|_p$ is neither a redex in $\mathcal{U}$ nor a variable bound by a redex in $\mathcal{U}$, then for some $i \in \pos{s|_p} \cap \natnum$ the node $n'$ is labelled $(s, p \cdot i)$ and the edge from $n$ to $n'$ is labelled~$i$,
\item
if $s|_p$ is a redex $u \in \mathcal{U}$ with $l \rew r$ the employed rewrite rule, then the node $n'$ is labelled $(r, \epsilon, \redpos{u})$ and the edge from $n$ to $n'$ is unlabelled,
\item
if $s|_p$ is a variable $x$ bound by a redex $u \in \mathcal{U}$ with $l \rew r$ the employed rewrite rule, then the node $n'$ is labelled $(r, p' \cdot i, \redpos{u})$ and the edge from $n$ to $n'$ is unlabelled, such that $(r, p', \redpos{u})$ was the last node before $n$ with $\redpos{u}$, $\rs{r|_{p'}} = Z$, and $l|_{q \cdot i} = x$ with $q$ the unique position of $Z$ in $l$.
\end{enumerate}
If a node $n$ of a path is labelled $(r, p, \redpos{u})$ and has an outgoing edge to a node $n'$, then:
\begin{enumerate}[(1)]
\item
if $\rs{r|_p}$ is not a meta-variable, then for some $i \in \pos{r|_p} \cap \natnum$ the node $n'$ is labelled $(r, p \cdot i, \redpos{u})$ and the edge from $n$ to $n'$ is labelled $i$,
\item
if $\rs{r|_p}$ is a meta-variable $Z$, then the node $n'$ is labelled $(s, \redpos{u} \cdot q)$ and the edge from $n$ to $n'$ is unlabelled, such that $l \rew r$ is the rewrite rule employed in $u$ and such that $q$ is the unique position of $Z$ in $l$.
\end{enumerate}
\end{defi}

A path ends in case we encounter a nullary function symbol or a variable not bound by a redex of $s$ in $\mathcal{U}$ (this is automatically the case for any variable that occurs on the right-hand side of a rewrite rule). This is immediate by the fact that $\pos{t} \cap \natnum$ is empty in case $t$ is either a variable or nullary function symbol.

We say that a path is \emph{maximal} if it is not a proper prefix of another path. We write a path $\apath$ as a (possibly infinite) sequence of alternating nodes and edges $\apath = n_1 e_1 n_2 \cdots$. 

\begin{defi}
Let $\apath = n_1 e_1 n_2 \cdots $ be a path of $s$ with respect to $\mathcal{U}$. The \emph{path projection} $\phi(\apath)$ of $\apath$ is a sequence of alternating nodes and edges $\phi(\apath) = \phi(n_1) \phi(e_1) \phi(n_2) \cdots$. Each node $\phi(n)$ is either unlabelled or labelled with a function symbol or variable such that:
\begin{enumerate}[(1)]
\item
if $n$ is labelled $(s, p)$, then $\phi(n)$ is unlabelled if $s|_p$ is a redex in $\mathcal{U}$ or a variable bound by such a redex and it is labelled $\rs{s|_p}$ otherwise, and
\item
if $n$ is labelled $(r, p, q)$, then $\phi(n)$ is unlabelled if $\rs{r|_p}$ is a meta-variable and it is labelled $\rs{r|_p}$ otherwise.
\end{enumerate}
Each edge $\phi(e)$ is either labelled with an element of $\natnum$ or labelled $\epsilon$ such that if $e$ is labelled $i$, then $\phi(e)$ has the same label, and if $e$ is unlabelled, then $\phi(e)$ is labelled $\epsilon$.
\end{defi}

\noindent Note that the nodes of path projections are either unlabelled or labelled with \emph{function symbols} or \emph{variables}; this is contrary to paths whose nodes are labelled with \emph{pairs} and \emph{triples}. 

\begin{exa}
\label{ex:paths}
Consider the orthogonal iCRS that only has the following rewrite rule, also denoted $l \rew r$:
\[
f([x]Z(x), Z') \rew Z(g(Z(Z'))) \, .
\]
Given the terms $s = f([x]g(x), a)$ and $t = g(g(g(a)))$ and the set $\mathcal{U}$ containing the only redex in $s$, we have that $s \rew t$ is a complete development of $\mathcal{U}$.

The term $s$ has one maximal path with respect to $\mathcal{U}$:
\begin{multline*}
(s, \epsilon)
  \rightarrow (r, \epsilon, \epsilon)
  \rightarrow (s, 10)
  \overset{1}{\rightarrow} (s, 101)
  \rightarrow (r, 1, \epsilon) 
  \overset{1}{\rightarrow} (r, 11, \epsilon)  \\
\rightarrow (s, 10)
  \overset{1}{\rightarrow} (s, 101)
  \rightarrow (r, 111, \epsilon)
  \rightarrow (s, 2)
\end{multline*}
Moreover, the term $t$ has one maximal path with respect to $\mathcal{U}/(s \rew t) = \emptyset$:
\[
(t, \epsilon)
  \overset{1}{\rightarrow} (t, 1)
  \overset{1}{\rightarrow} (t, 11)
  \overset{1}{\rightarrow} (t, 111) \, .
\]
The path projections of the maximal paths are, respectively,
\[
\cdot
  \overset{\epsilon}{\rightarrow} \cdot
  \overset{\epsilon}{\rightarrow} g
  \overset{1}{\rightarrow} \cdot
  \overset{\epsilon}{\rightarrow} g
  \overset{1}{\rightarrow} \cdot
  \overset{\epsilon}{\rightarrow} g
  \overset{1}{\rightarrow} \cdot
  \overset{\epsilon}{\rightarrow} \cdot
  \overset{\epsilon}{\rightarrow} a
\]
and
\[
g
  \overset{1}{\rightarrow} g
  \overset{1}{\rightarrow} g
  \overset{1}{\rightarrow} a \, .
\]
\end{exa}

Let $\mathcal{P}(s, \mathcal{U})$ denote the set of path projections
of \emph{maximal paths} of $s$ with respect to $\mathcal{U}$. The
following two results from \cite{JJ05a} can be witnessed in the above example.
The proof of the first can be found in Appendix \ref{app:proof_III}.

\begin{prop}
\label{prop:phibiject}
The map $\phi$ defines a bijection between the set of paths and the set
of path projections, respectively between maximal paths and the path
projections in $\mathcal{P}(s, \mathcal{U})$.
\end{prop}

\begin{lem}
\label{unlabeldel}
Let $u \in \mathcal{U}$ and let $s \rew t$ be the rewrite step contracting $u$. There exists a bijection between $\mathcal{P}(s, \mathcal{U})$ and $\mathcal{P}(t, \mathcal{U}/u)$. Given a path projection $\phi(\apath) \in \mathcal{P}(s, \mathcal{U})$, its image under the bijection is obtained by deleting finite sequences of unlabelled nodes and $\epsilon$-labelled edges from $\phi(\apath)$. \qed
\end{lem}

We continue with the definition of the finite jumps property, a property of $\mathcal{U}$ depending on $\mathcal{P}(s, \mathcal{U})$. We also introduce some terminology to relate a term to $\mathcal{P}(s, \mathcal{U})$.

\begin{defi}
The set $\mathcal{U}$ has the \emph{finite jumps property} if no path projection occurring in $\mathcal{P}(s, \mathcal{U})$ contains an infinite sequence of unlabelled nodes and $\epsilon$-labelled edges. Moreover, a term $t$ \emph{matches} $\mathcal{P}(s, \mathcal{U})$ if, for all $\phi(\apath) \in \mathcal{P}(s, \mathcal{U})$ and all prefixes of $\phi(\apath)$ ending in a node $\phi(n)$ labelled $f$, it holds that $\rs{t|_p} = f$, where $p$ is the concatenation of the edge labels in the prefix (starting at the first node of $\phi(\apath)$ and ending in $\phi(n)$).
\end{defi}

With respect to the finite jumps property the following three results are proven in \cite{JJ05a}.

\begin{prop}
\label{uniqueterm}
If $\mathcal{U}$ has the finite jumps property, then there exists a unique term, denoted $\mathcal{T}(s, \mathcal{U})$, that matches $\mathcal{P}(s, \mathcal{U})$. \qed
\end{prop}

\begin{thm}[Finite Jumps Developments Theorem]
\label{fjdt}
If $\mathcal{U}$ has the finite jumps property, then:
\begin{enumerate}[\em(1)]
\item
every complete development of $\mathcal{U}$ ends in $\mathcal{T}(s, \mathcal{U})$,
\item
for any $p \in \pos{s}$, the set of descendants of $p$ by a complete development of $\mathcal{U}$ is independent of the complete development,
\item
for any redex $u$ of $s$, the set of residuals of $u$ by a complete development of $\mathcal{U}$ is independent of the complete development, and
\item
$\mathcal{U}$ has a complete development. \qed
\end{enumerate}
\end{thm}

\begin{lem}
\label{compimplfinite}
The set $\mathcal{U}$ has a complete development if{f} $\mathcal{U}$ has the finite jumps property. \qed
\end{lem}

\begin{rem}
\label{rem:label}
The proof of Theorem \ref{fjdt}(2) is based on a labelling. In analogy to \cite[Section II.2]{K80}, it presupposes a set of labels $\mathcal{K}$ including a special \emph{empty} label $\varepsilon$. Using the labels, \emph{labelled alternatives} are defined for all function symbols $f$ and variables $x$ and for all labels $k \in \mathcal{K}$, these are denoted $f^k$ and $x^k$, where $f$ and $f^k$ have the same arity. A \emph{labelling} of a (meta-)term replaces each function symbol and variable (including the variables that occur in abstractions) by a labelled alternative, assuming that the labels of variables are ignored where \emph{bindings} and \emph{valuations} are concerned.

The labelled version of the assumed orthogonal iCRS includes for every rewrite rule $l \rew r$ and every possible labelling $l'$ of $l$ a rewrite rule $l' \rew r'$, where $r'$ is the labelling of $r$ that labels all function symbols and variables with $\varepsilon$. The labelled version of the iCRS is easily shown to be orthogonal (see \cite[Proposition II.2.6]{K80}).

Each reduction in the labelled version corresponds to a reduction in the original iCRS by removal of all labels. Moreover, given a reduction in the original iCRS and a labelling for the initial term, there exists a unique reduction in the labelled version such that removal of the labels gives the reduction we started out with. Finally, given a term in which some subterms are labelled $k$, the descendants of these subterms across some reduction are precisely the subterms labelled $k$ in the final term. These descendants are exactly the descendants obtained in the corresponding unlabelled reduction.
\end{rem}

\section{Overview and roadmap}
\label{sec:overview}

From this point onwards, we concentrate on the exposition and development of new results, assuming fully-extended, orthogonal iCRSs throughout. The present section provides a high-level overview of the novel proof technique employed and a roadmap to the results.

\subsection{Overview of the proof technique}
\label{sec:overoverview}

We start with an overview of the employed proof technique, the proper technical development of which begins in Section \ref{sec:ess}. The technique is a variant of van Oostrom's technique of essential rewrite steps \cite{O99} as developed for finitary higher-order systems.

Two observations are important to understand the employed technique. First, in both (finitary) term rewriting and first-order infinitary rewriting the use of projections is the fulcrum of most proofs (that we know of) concerning reduction strategies and confluence. This is effectively an application of the Strip Lemma, which states that a reduction can be projected over a single rewrite step. Unfortunately, the Strip Lemma fails in the infinitary higher-order case, as can already be witnessed in \iLC \cite{KKSV97,T03_KV}.

Second, both in the case of reduction strategies and confluence it is possible to limit our attention to finite parts of terms. For reduction strategies this requires one of the basic techniques from infinitary rewriting: considering terms --- in this case normal forms --- up to a certain finite depth for increasingly greater depths. In the case of confluence we are interested in redexes and as we assume fully-extended, orthogonal systems, it suffices to consider the redex patterns of these redexes, and those patterns are finite.

As the Strip Lemma does not hold for iCRSs, most proofs regarding strategies and confluence cannot be redeployed directly in the context of iCRSs. However, as we can limit our attention to finite parts of terms, it follows by strong convergence that only a finite number of steps along a reduction can actually `contribute' to function symbols that occur in a certain finite part of a term under consideration.

The key idea of the technique is now to `filter' the rewrite steps along a reduction based on their contribution to a chosen finite part of the final term of the reduction; keeping the steps that contribute --- the \emph{essential} steps --- and discarding the ones that do not --- the \emph{inessential} steps. This yields a finite reduction which is identical to the reduction being filtered as far as essential steps and the finite part under consideration are concerned. Since the reduction is finite, it can be projected over rewrite steps starting in the first term of the reduction (by repeated application of Proposition \ref{prop:finite_over_normal}).

The combination of filtering and projecting defines a new kind of projection in the sense that given a reduction and a rewrite step starting in the first term of the reduction a new reduction is obtained. Given this new kind of projection it becomes possible to redeploy the first-order technique of Sekar and Ramakrishnan \cite{SR90} in the context of fully-extended, orthogonal iCRSs. The technique is essentially a termination argument: A measure on reductions is defined which decreases when the reductions are projected across certain rewrite steps.

In the technical development below, instead of finite reductions, we consider finite sequences of complete developments, i.e.\ reductions consisting of a finite number of such developments. We are forced to do this as projecting a single rewrite step might actually yield an infinite number of residuals. In addition, we need a particular analysis of which positions depend on which other positions across rewrite steps that goes beyond ordinary descendant tracking; we call this analysis `propagation'. It will allow us to establish which rewrite steps are essential and which ones are not.

We continue to provide some more details regarding the three technical ingredients of the technique: propagation, measure, and projection. 

\subsubsection*{Propagation.}

As mentioned above, we are interested in either the part of a term up to a certain depth or a redex pattern. In both cases it holds for the positions that occur that all their prefix positions also occur. This leads us to only consider the propagation of so-called \emph{prefix sets}. These are \emph{finite} sets of positions such that if a position is included in the set, then all its prefix positions are also included in the set.

The propagation of prefix sets through finite sequences of complete developments now takes the form of a map $\pmap$. The map is first defined on complete developments $s \dev^\mathcal{U} t$: Given a prefix set $P$ of $t$ the map yields a prefix set of $s$, which we denote by $\pmapp{P}{s \dev^\mathcal{U} t}$. Intuitively, a position occurs in $\pmapp{P}{s \dev^\mathcal{U} t}$ if it `contributes' to the prefix set of $t$. Moreover, we call a position $p$ of $s$, respectively a redex $u$ in $s$, \emph{essential} if $p$, respectively the position of $u$, occurs in $\pmapp{P}{s \dev^\mathcal{U} t}$. They are called \emph{inessential} otherwise.

Using the fact that the set of positions obtained through application of $\pmap$ is a prefix set, both $\pmap$ and the notion of (in)essentiality are easily extended inductively to finite sequences of complete developments.

\subsubsection*{Measure.} 

The measure, which we denote by $\pme$, assigns to each finite sequence of complete developments and prefix set of its final term a tuple of the same length as the finite sequence. Each element of the tuple, which is natural number, effectively represents the number of essential steps in one of the developments of the finite sequence of complete developments. Tuples are compared first length-wise and next lexicographically (in the natural order). This yields a well-founded order --- as the natural order on natural numbers is well-founded --- which we denote by $\prec$.

\subsubsection*{Projection.}

The new kind of projection, called the \emph{emaciated projection} and denoted $\emmy$, projects finite sequences of complete developments over rewrite steps and is parametric in the prefix set assumed for the final term of such a sequence.

Given a redex $u$ in the first term of the finite sequence of developments $D$ considered, the emaciated projection behaves according to the essentiality of $u$. To be precise, given a prefix set $P$ in the final term of $D$, we will show that:

\begin{enumerate}[(1)]
\item
if $u$ is essential and no residual from $u/D$ occurs in $P$, then the projection yields a finite sequence of complete developments $D' = D \emmy u$ such that we have $\pmep{P}{D'} \prec \pmep{P}{D}$, and
\item
if $u$ is inessential, then the projection yields a finite sequence of complete developments $D' = D \emmy u$ such that we have $\pmep{P}{D'} = \pmep{P}{D}$ and $\pmapp{P}{D'} = \pmapp{P}{D}$.
\end{enumerate}

In both cases, $D'$ is of the same length as $D$, starting in the term created by $u$, such that the function symbols in final terms of $D$ and $D'$ are identical as far as positions in $P$ are concerned.

The case where $u$ is essential facilitates the termination argument mentioned  above. The case where $u$ is inessential shows that only the positions in $\pmapp{P}{D}$ `contribute' to the prefix set of the final term of $D$, as mentioned above while introducing $\pmap$.

The case in which $u$ is essential and in which some residual from $u/D$ does occur in $P$, i.e.\ the only case not covered by the above clauses, will be dealt with by the technical provision that tuples are first compared length-wise.

\begin{rem}
As can be inferred from the above, van Oostrom's technique of essential rewrite steps \cite{O99}, as adapted by us, incorporates a proof technique originally developed by Sekar and Ramakrishnan \cite{SR90} to study normalising strategies in first-order rewriting; a technique later refined by Middeldorp \cite{M97}. In fact, van Oostrom's technique can be seen as a higher-order variant of the techniques by Sekar and Ramakrishnan and Middeldorp. Unlike van Oostrom, the latter do not require the introduction of the notion of essentiality, which derives from \cite{K88,GK94}.

The filtering described earlier in this section already makes sense in the finite higher-order case, as dealt with by van Oostrom \cite{O99}: It also helps to cope with the nestings that can occur in these systems when defining the appropriate measure.

Contrary to all other techniques, which apply only to finite reductions, our instalment revolves around finite sequences of complete developments. As stated previously, the shift to finite sequences is \emph{necessary} in the setting of infinitary rewriting because projecting one reduction step over another may yield an infinite complete development of the residuals of the projected redex.
\end{rem}

\subsection{Roadmap to the results}

The main results of this paper are Theorems \ref{thm:outermostfair}, \ref{thm:fair}, and \ref{thm:neededfair}, showing, respectively, that the outermost-fair, fair, and needed-fair reduction strategies are normalising for orthogonal, fully-extended iCRSs.

\begin{figure}
\[
\underbrace{
\xymatrix@=0.3cm{
\txt{Definitions \ref{def:pmap} and \ref{def:essential}\\(Essentiality)} \ar@{=>}[dd] \ar@{=>}[ddr] & & \txt{Definition \ref{measuredef}\\(Measure)} \ar@{=>}[ddl] \\
\\
\txt{Proposition \ref{prop:desess}} & \txt{Lemma \ref{lem:simple_proj}} \ar@{=>}[dd] \\
\\
& \txt{Definition \ref{def:emmy}\\ \makebox[0pt][c]{(Emaciated projection across $\rew$)}}
}
}_{\text{\large$\Downarrow$}}
\]

\[
\underbrace{
\xymatrix@=0.3cm{
\txt{Lemma \ref{lem:essential_then_decrease}} \ar@{=>}[ddr] & & \txt{Lemma \ref{lem:inessential_nonroot_then_equal}} \ar@{=>}[ddl] \\
\\
& \txt{Definition \ref{def:emmybig}\\ \makebox[0pt][c]{(Emaciated projection across $\trewt$)}}
}
}_{\text{\large$\Downarrow$}}
\]
\[
\xymatrix@=0.3cm{
& \txt{Lemma \ref{lem:devsplit} \\ Corollary \ref{cor:devsplit}} \ar@{=>}[dd] & \txt{Lemma \ref{lem:nf_mirror}\\(Uses Lemma \ref{lem:simple_proj})} \ar@{=>}@/^5.0pc/[dddd]\\
\\
& \txt{Lemma \ref{lem:essstaysess_prim}\\(Uses Proposition \ref{prop:desess})} \ar@{=>}[dd] & \txt{Lemma \ref{lem:nf_submirror}\\(Uses Proposition \ref{prop:desess})} \ar@{=>}[dd] \\
\\
 & \txt{Lemma \ref{lem:essstaysess}\\(Uses Lemma \ref{lem:simple_proj})} \ar@{=>}[dd] & \txt{Lemma \ref{lem:nf_ok}} \ar@{=>}[dd] \ar@{=>}[ddl] \\
\ar@{.}[rrr] & & & \\
 & \txt{Theorem \ref{thm:neededfair}\\(Uses Lemmas \ref{lem:essential_then_decrease} and \ref{lem:inessential_nonroot_then_equal})} & \txt{Theorems \ref{thm:outermostfair} and \ref{thm:fair}\\(Uses Lemmas \ref{lem:simple_proj}, \ref{lem:essential_then_decrease}, and \ref{lem:inessential_nonroot_then_equal})}
}
\]
\caption{\label{fig:road}Roadmap to the results}
\end{figure}
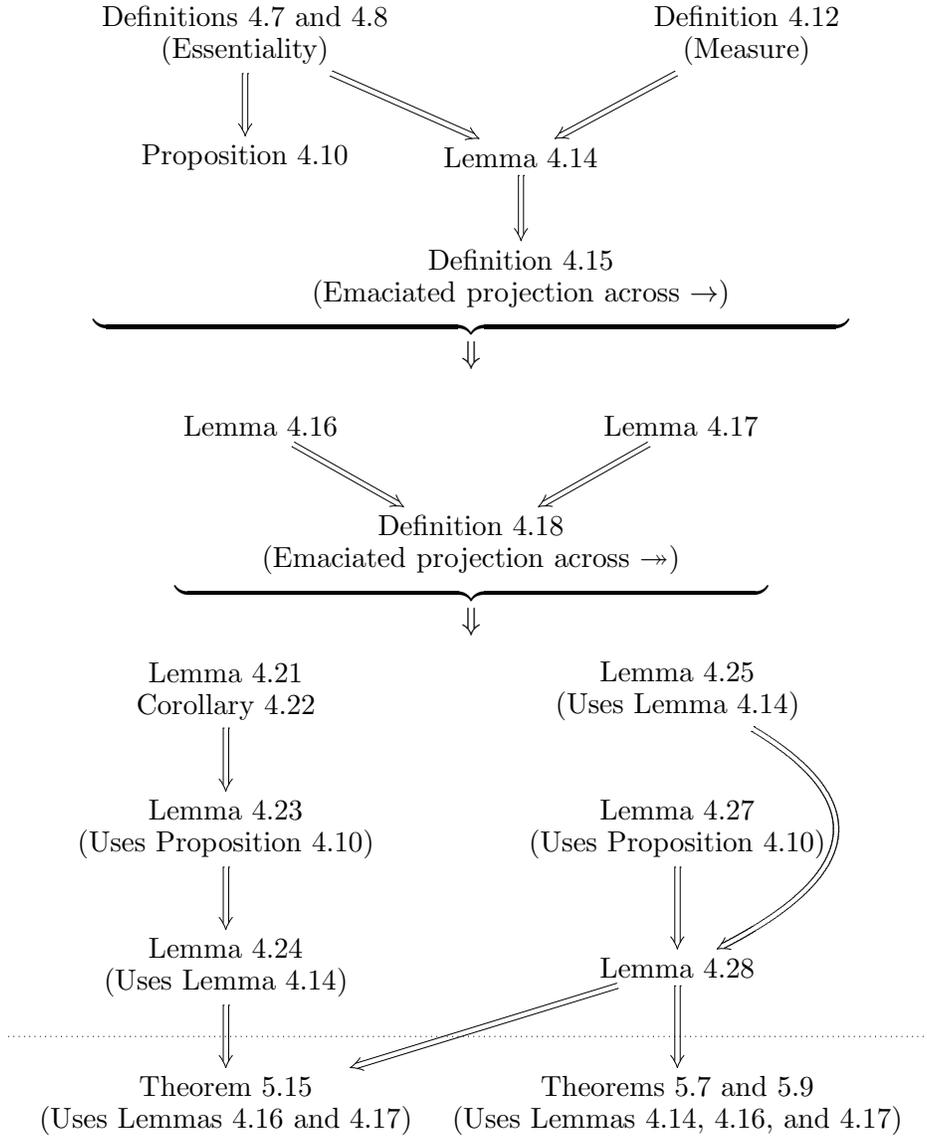

Up to the proofs of the main results, the paper can be divided into three parts (see Figure \ref{fig:road}). The first part, formed by Section \ref{sec:prop_pre} and the first half of Section \ref{sec:mes_pro}, introduces the elements of the proof technique as discussed above. In particular, Proposition \ref{prop:desess} states that the map $\pmap$ behaves as expected and Lemma \ref{lem:simple_proj} shows that emaciated projections indeed project finite sequences of complete developments.

The second part, taking up the second half of Section \ref{sec:mes_pro}, shows that emaciated projections satisfy the properties stated above (Lemmas \ref{lem:essential_then_decrease} and \ref{lem:inessential_nonroot_then_equal}). The part also extends the concept of emaciated projections from projections across rewrite steps to projections across reductions (Definition \ref{def:emmybig}).

The third part, formed by Sections \ref{sec:ess_prop} and \ref{sec:nf_emmy}, establishes some further relations between essential redexes on the one side and complete developments and projections on the other (Lemma \ref{lem:essstaysess}). In addition, the third part relates emaciated projections and reductions to normal form (Lemma \ref{lem:nf_ok}).

\section{Essential rewrite steps}
\label{sec:ess}

We now proceed as outlined above. In Section \ref{sec:prop_pre} we define the map $\pmap$ on prefixes and complete developments. Thereafter, in Section \ref{sec:mes_pro} the measure and emaciated projection are formally introduced and it is shown that the projection satisfies the aforementioned properties. In Section \ref{sec:ess_prop} we prove some further properties of essential positions and redexes with regard to complete developments and projections. Finally, in Section \ref{sec:nf_emmy} we relate emaciated projections and reductions to normal form.

Before we begin our development, we formally define prefix sets:
\begin{defi}
A \emph{prefix set} of a term $s$ is a finite set $P \subseteq \pos{s}$ such that all prefixes of positions in $P$ are also in $P$.
\end{defi}

\noindent
Take heed that prefix sets are \emph{finite}!

\begin{rem}
The material in this section completely redevelops the theory of essential rewrite steps for iCRSs that previously appeared in \cite{JJ05b}. The current theory allows for rewrite rules with infinite right-hand sides, while previously only finite right-hand sides were allowed.
\end{rem}

\subsection{Propagation of prefix sets}
\label{sec:prop_pre}

To define the propagation of prefix sets over complete developments, we relate prefix sets with paths, whose definition can be found in Section \ref{sec:pathsjumps}. In particular, we employ the notion of a \emph{path prefix set} which includes paths that `occur' in a prefix set of a term where some set of redexes is present. Using the path prefix sets we recover the positions `encountered' when defining the paths in these sets, in particular the positions of the redex patterns encountered.

\begin{defi}
Let $s$ and $t$ be terms, $\mathcal{U}$ a set of redexes in $s$ such that $s \dev^\mathcal{U} t$ and $P$ a prefix set of $t$. The \emph{path prefix set} of $P$ with respect to $\mathcal{U}$ is the set of all paths $\apath$ of $s$ with respect to $\mathcal{U}$ such that the concatenation of the edge labels of the path projection $\phi(\apath)$ is in $P$.
\end{defi}

Observe that if a certain path is included in a path prefix set, then all its prefixes are also included in the set. This follows by the dependence on prefix sets and their closure under the prefix relation on positions.

\begin{exa}
\label{ex:path_prefixes}
Consider the iCRS from Example \ref{ex:paths}, where
$s = f([x]g(x),a)$ and $t = g(g(g(a)))$.
The set $P = \{ \epsilon, 1, 11 \}$ is a prefix set of $t$.
Let $\mathcal{U}$ be the set containing the only redex of $s$ and
observe that $s \dev^\mathcal{U} t$.
The path prefix set of $P$ with respect to $\mathcal{U}$ is the set of all paths
that are prefixes of
\[
(s,\epsilon) \rew (r,\epsilon,\epsilon) \rew (s,10)
\overset{1}{\rew} (s,101) \rew (r,1,\epsilon) \overset{1}{\rew}
(r,11,\epsilon) \rew (s,10) \, .
\]
\end{exa}

To recover the positions `encountered' when defining the paths in path prefix sets, we use the following map.

\begin{defi}
Let $s$ be a term and $\mathcal{U}$ a set of redexes in $s$. The map $\zeta$ from \emph{finite} paths $\apath$ of $s$ with respect to $\mathcal{U}$, with final node $n$, to \emph{finite} subsets of $\pos{s}$ is defined as follows:
\[
\zeta(\apath) =
\begin{cases}
\{ p \}   & \text{if $n = (s, p)$ and no redex
                  in $\mathcal{U}$ occurs at $p$} \\
Q         & \text{if $n = (s, p)$ and a redex $u \in \mathcal{U}$ occurs
                  at $p$} \\
\emptyset & \text{if $n = (r, p, \redpos{u})$}
\end{cases}
\]
where $Q$ is the set of positions of $s$ that occur in the redex pattern of $u$.
\end{defi}

The following lemma shows that $\zeta$ can be extended to a well-defined function on path prefix sets yielding a prefix set:

\begin{lem}
\label{lem:zetaprefix}
Let $s$ and $t$ be terms, $\mathcal{U}$ a set of redexes in $s$ such that $s \dev^\mathcal{U} t$, and $P$ a prefix set of $t$. If $\Psi$ is the path prefix set of $P$ with respect to $\mathcal{U}$, then $\zeta(\Psi) = \{ \zeta(\apath) \seper \apath \in \Psi\}$ is well-defined and yields a prefix set of $s$.
\end{lem}

\proof
Let $\Psi$ be the path prefix set of $P$ with respect to $\mathcal{U}$. Since $\mathcal{U}$ has a complete development, it follows by Lemma \ref{compimplfinite} that $\mathcal{U}$ has the finite jumps property, i.e.\ all path projections in $\mathcal{P}(s, \mathcal{U})$ contain only finite sequences of unlabelled nodes and $\epsilon$-labelled edges. As each path is a prefix of a maximal path, whose path projections are in $\mathcal{P}(s, \mathcal{U})$, it follows by definition of path projections and the finite jumps property that each path in $\Psi$ is finite. Hence, $\zeta(\Psi)$ is well-defined.

This leaves to show that $\zeta(\Psi)$ yields a prefix set of $s$, i.e.\ that $\zeta(\Psi)$ is finite and that each prefix of a position in $\zeta(\Psi)$ is also in $\zeta(\Psi)$.

For each position in the prefix set $P$ of $t$ a finite number of paths is included in $\Psi$. This follows by induction on the length of the positions, employing the fact that $\mathcal{U}$ has the finite jumps property and the fact that the extension of a path is uniquely determined by the definition of paths and the considered position. Since $P$ is finite, the same now follows for $\Psi$. Hence, as $\zeta$ maps each finite path to a finite number of positions, $\zeta(\Psi)$ is finite.

Let $p \in \zeta(\Psi)$ and $q < p$. There are two possibilities: $q$ occurs either in the redex pattern of a redex in $\mathcal{U}$, or not. In case $q$ occurs in the redex pattern of a redex $u \in \mathcal{U}$, it follows by $p \in \zeta(\Psi)$ and the definition of paths that there exists a path in the path prefix set which ends in the node $(s, \redpos{u})$, with $\redpos{u}$ the position of the redex $u$. Hence, in this case $q \in \zeta(\Psi)$ by definition of $\zeta$. In case $q$ does not occur in a redex pattern, it follows by the definition of paths and the inclusion of $p$ in $\zeta(\Psi)$ that there exists a path in the path prefix set which ends in the node $(s, q)$ and, thus, $q \in \zeta(\Psi)$. Hence, all prefixes of positions in $\zeta(\Psi)$ are included in $\zeta(\Psi)$. Employing the finiteness of $\zeta(\Psi)$ it now follows that $\zeta(\Psi)$ yields a prefix of $s$, as required.
\qed

By the previous lemma, the map $\pmap$ shortly described in the previous section can now be defined as follows:
\begin{defi}
\label{def:pmap}
Let $s$ and $t$ be terms and $\mathcal{U}$ a set of redexes in $s$ such that $s \dev^\mathcal{U} t$. The map $\pmap$ from prefix sets $P$ of $t$ to prefix sets of $s$ is defined as:
\[
\pmapp{P}{s \dev^\mathcal{U} t} = \zeta(\Psi)
\]
where $\Psi$ is the path prefix set of $P$ with respect to $\mathcal{U}$.
\end{defi}

The following definition will be useful in the context of the map $\pmap$ and gives the name to the proof technique being introduced.
\begin{defi}
\label{def:essential}
Let $s$ and $t$ be terms, $\mathcal{U}$ a set of redexes in $s$ such that $s \dev^\mathcal{U} t$, and $P$ a prefix set of $t$. A position $p$ of  $s$, respectively a redex $u$ in $s$, is called \emph{essential} for $P$ if $p$, respectively the position of $u$, occurs in $\pmapp{P}{s \dev^\mathcal{U} t}$. A position, respectively a redex, is called \emph{inessential} otherwise.
\end{defi}

\begin{exa}
Consider the prefix set $P$ in Example \ref{ex:path_prefixes}. We have that the positions $\epsilon$, $1$, $10$, and $101$ are essential for $P$ in $s$.
\end{exa}

As the set of positions obtained through application of $\pmap$ is a prefix set, the map is easily extended to a finite sequence of complete developments $s_0 \dev^{\mathcal{U}_1} s_1 \dev^{\mathcal{U}_2} \cdots \dev^{\mathcal{U}_n} s_n$: In case of $s_n$, define $\pmaph{P}$ to be $P$. In case of $s_i$, with $i < n$, define $\pmaph{P}$ to be $\pmapp{P_{i + 1}}{s_i \dev^{\mathcal{U}_{i + 1}} s_{i + 1}}$ where $P_{i + 1}$ is the prefix set obtained for $s_{i + 1}$. The notion of (in)essentiality is extended accordingly.

To end this section, we show that an essential position will always descend to a position in the assumed prefix set in case the position does not occur in the redex pattern of a redex in the assumed complete development.
\begin{prop}
\label{prop:desess}
Let $s$ and $t$ be terms, $\mathcal{U}$ a set of redexes in $s$ such that $s \dev^\mathcal{U} t$, and $P$ a prefix set of $t$. If $p \in \pos{s}$ does not occur in the redex pattern of a redex in $\mathcal{U}$ and is not the position of a variable bound by a redex in $\mathcal{U}$, then $p$ is essential if{f} there exists a position $q \in P$ such that $q \in p/(s \dev t)$ and $p$ is inessential if{f} no descendant of $p$ occurs in~$P$.
\end{prop}

\proof
By Lemma \ref{compimplfinite}, it follows that $\mathcal{U}$ has the finite jumps property. Employing the labelling from the proof of Theorem \ref{fjdt}(2) and its properties relating labelled and unlabelled reductions and descendants across these reductions --- as exhibited in Remark \ref{rem:label} --- together with Theorem \ref{fjdt}(1), it is easy to see that a position $p \in \pos{s}$ descends to a position $q \in \pos{t}$ if{f} $p$ does not occur in a redex pattern of a redex in $\mathcal{U}$ and there exists a finite path $\apath$ with final node $n = (s, p)$ such that $\phi(n)$ is labelled and such that the concatenation of the edge labels of the path projection of $\apath$ is $q$. Since $\zeta(\apath) = \{ p \}$, the result follows by definition of path prefix sets.
\qed

\subsection{Measure and projection}
\label{sec:mes_pro}

In this section, we define the measure on finite sequences of complete developments and the emaciated projection. To facilitate our exposition, we fix the following notation with regard to sequences of complete developments.

\begin{notation}
By $D$, respectively $E$, we denote a finite sequence of complete developments $\fscd{s}{U}{n}$, respectively $\fscd{t}{V}{n}$, of length $n$. Moreover, if $P$ is a prefix set of $s_n$, then for all $0 \leq i \leq n$ we denote by $P_i$ the set of positions essential for $P$ in $s_i$.
\end{notation}

We define the measure on finite sequences of complete developments with respect to prefix sets:
\begin{defi}
\label{measuredef}
The \emph{measure} $\pmep{P}{D}$ of $D$ with respect to the prefix set $P$ of $s_n$ is the $n$-tuple $(l_n,\ldots,l_1)$ --- note the reverse order! --- such that $l_i$, with $1 \leq i \leq n$, is the cardinality of the path prefix set of $P_i$ with respect to $\mathcal{U}_i$.
\end{defi}
As already mentioned, the tuples in the above definition are compared first length-wise and next lexicographically (in the natural order). This yields a well-founded order, as each element of a tuple is finite by Lemma \ref{lem:zetaprefix}. We denote this order by $\prec$.

Before we continue with the definition of the emaciated projection, we define an auxiliary notion regarding prefix sets on the one hand and terms and finite sequences of complete developments on the other.
\begin{defi}
Let $s$ and $t$ be terms and $P$ a prefix set of $s$. The term $t$ \emph{mirrors} $s$ in $P$, if $P \subseteq \pos{t}$ and $\rs{t|_p} = \rs{s|_p}$ for all $p \in P$.

Let $P$ be a prefix set of $s_n$ in $D$. The finite sequence $E$ \emph{mirrors} $D$ in $P$ if for all $0 \leq i \leq n$ it holds that the set of positions essential for $P$ in $t_i$ is $P_i$, $t_i$ mirrors $s_i$ in $P_i$, and the path prefix set of $P_i$ with respect to $\mathcal{V}_i$ is identical to the path prefix set of $P_i$ with respect to $\mathcal{U}_i$.
\end{defi}

The following lemma is key in the definition of the emaciated projection:
\begin{lem}
\label{lem:simple_proj}
Let $P$ be a prefix set of $s_n$ in $D$ and let $t_0$ mirror $s_0$ in the positions essential for $P$ in $s_0$. There exists a finite sequence $E$, with $\mathcal{V}_i$ for all $1 \leq i \leq n$ \emph{finite} and consisting \emph{solely} of essential redexes, such that $E$ mirrors $D$ in $P$ and $\pmep{P}{E} = \pmep{P}{D}$.
\end{lem}

\proof
By induction on $n$, the number of complete developments in $D$. In case $n = 0$, the result is immediate by definition of $t_0$.

In case $n > 0$, let $\mathcal{U}'_n$ contain the redexes from $\mathcal{U}_n$ essential for $P$. Observe for each $u \in \mathcal{U}'_n$ that all positions in the redex pattern of $u$ occur at positions in $P_{n - 1}$ by definition of the map $\zeta$. Hence, since we have by the induction hypothesis that $t_{n - 1}$ mirrors $s_{n - 1}$ in $P_{n - 1}$, it follows by orthogonality and fully-extendedness that there exists for each redex in $\mathcal{U}'_n$ a redex in $t_{n - 1}$ at the same position and employing the same rewrite rule. Define $\mathcal{V}_n$ to be the set of these corresponding redexes in $t_{n - 1}$. Obviously, the sets $\mathcal{V}_n$ and $\mathcal{U}'_n$ have the same cardinality, which is finite as $P_{n - 1}$ is finite.

Since $\mathcal{V}_n$ is finite, it follows by Lemma \ref{lem:finite_dev} that there exists a complete development $t_{n-1} \dev^{\mathcal{V}_n} t_n$. Moreover, since $P_{n - 1}$ is a prefix set and $t_{n - 1}$ mirrors $s_{n - 1}$ in $P_{n - 1}$, it follows by definition of paths and $\mathcal{V}_n$ that for each path of $s_{n - 1}$ with respect to $\mathcal{U}_n$ occurring in the path prefix set of $P$ there exists an identical path of $t_{n - 1}$ with respect to $\mathcal{V}_n$. Hence, by definition of path projections, we have for the terms matching $\mathcal{P}(s_{n - 1}, \mathcal{U}_n)$ and $\mathcal{P}(t_{n - 1}, \mathcal{V}_n)$, i.e.\ $s_n$ and $t_n$, that $P \subseteq \pos{t_n}$, $\rs{s_n|_p} = \rs{t_n|_p}$ for all $p \in P$, and that all positions in $P_{n - 1}$ and redexes in $\mathcal{V}_n$ are essential for $P$. The induction hypothesis now furnishes the result.
\qed

Observe that the above lemma `cuts down' the sets of redexes that occur in the sequence of complete developments to \emph{finite} sets consisting solely of essential redexes. The lemma states that this suffices to obtain a term $t_n$ with prefix $P$.

We can now define our projection:
\begin{defi}
\label{def:emmy}
Let $P$ be a prefix set of $s_n$ in $D$. If $s_0 \rew t_0$ contracts a redex $u$ such that no redex in $u/D$ occurs at a position in $P$, then the \emph{emaciated projection} of $D$ across $s_0 \rew t_0$ with respect to $P$, written $D \emmy u$, is defined as $E/u$, where $E$ is the result of applying Lemma \ref{lem:simple_proj} to $D$ with $E$ starting in $s_0$.
\end{defi}

That the projection $E/u$ in the above definition exists follows by repeated application of Proposition \ref{prop:finite_over_normal}. The proposition can be applied since each set of redexes developed along $E$ is finite.

Since $E$ mirrors $D$ in $P$, orthogonality and fully-extendedness imply that no redex in $u/E$ occurs at a position in $P$ and, hence, the final term of $E/u$ mirrors the final one of $D$ in $P$. Moreover, in case $u$ is inessential, the requirement that no redex in $u/D$ occurs at a position in $P$ is void by Proposition \ref{prop:desess}. If such a redex would occur, it would be essential.

In the following two lemmas we relate the emaciated projection with the measure in the way discussed in Section \ref{sec:overview}.

\begin{lem}
\label{lem:essential_then_decrease}
Let $P$ be a prefix set of $s_n$ in $D$. If $s_0 \rew t_0$ contracts an essential redex $u$ such that no redex in $u/D$ occurs at a position in $P$, then $\pmep{P}{D \emmy u} \prec \pmep{P}{D}$.
\end{lem}

\proof
Suppose that $s_0 \rew t_0$ contracts an essential redex $u$ such that no redex in $u/D$ occurs at a position in $P$.  Denote by $E$ the result of applying Lemma \ref{lem:simple_proj} to $D$, with $E$ starting in $s_0$, and write $E/u$ as $\fscdp{t}{V}{n}$ where $\mathcal{V}'_i = \mathcal{V}_i / (t_{i - 1} \dev t'_{i - 1})$ for all $1 \leq i \leq n$.

Let $i < n$ be the largest index of a set $\mathcal{U}_i$ that contains a residual of $u$ that is essential. Since $E$ mirrors $D$ in $P$, the index $i$ is also the largest index of a set $\mathcal{V}_i$ that contains a residual of $u$ that is essential. No residual of $u$ occurs at an essential position in $t_j$ for $i < j \leq n$. Otherwise, a residual also occurs at an essential position in $t_{j + 1}$ by definition of residuals and Proposition \ref{prop:desess}. Iteratively, a residual then occurs in $t_n$ at a position in $P$, contradicting assumptions. Hence, by induction we have for all $i < j \leq n$ that $t'_j$ mirrors $t_j$ in $P_j$, where $P_j$ is the set of positions essential for $P$ in both $t'_j$ and $t_j$. Moreover, for each essential redex in $\mathcal{V}'_j$ there exists an essential redex in $\mathcal{V}_j$ at the same position and employing the same rewrite rule, and vice versa.

Write $\pmep{P}{E} = (l_n, \ldots, l_1)$ and $\pmep{P}{E/u} = (l'_n,\ldots,l'_1)$. For all $j < i$, the cardinality of the path prefix set of $\mathcal{V}'_j$ may be different from the one of $\mathcal{V}_j$, i.e.\ we may have $l'_j \not = l_j$. The cardinality of the path prefix set of $\mathcal{V}'_i$ is less than that of $\mathcal{V}_i$ by Proposition \ref{prop:phibiject} and Lemma \ref{unlabeldel}. Hence, $l'_i < l_i$. Finally, for all $i < k \leq n$ the path prefix sets of $\mathcal{V}'_k$ and $\mathcal{V}_k$ have the same cardinality by the correspondence between the essential redexes, i.e.\ $l'_k = l_k$. Thus, $\pmep{P}{E/u} \prec \pmep{P}{E}$. By Lemma \ref{lem:simple_proj} it now follows that $\pmep{P}{D \emmy u} \prec \pmep{P}{D}$, as required.
\qed

\begin{lem}
\label{lem:inessential_nonroot_then_equal}
Let $P$ be a prefix set of $s_n$ in $D$. If $s_0 \rew t_0$ contracts an inessential redex $u$, then $D \emmy u$ mirrors $D$ in $P$ and $\pmep{P}{D \emmy u} = \pmep{P}{D}$.
\end{lem}

\proof
Suppose that $s_0 \rew t_0$ contracts an inessential redex $u$. Denote by $E$ the result of applying Lemma \ref{lem:simple_proj} to $D$, with $E$ starting in $s_0$, and write $E/u$ as $\fscdp{t}{V}{n}$ where $\mathcal{V}'_i = \mathcal{V}_i / (t_{i - 1} \dev t'_{i - 1})$ for all $1 \leq i \leq n$.

For all $0 \leq i < n$, no residual of $u$ occurs at an essential position in $t_i$. Otherwise, as $E$ mirrors $D$ in $P$, it follows that $u$ is essential by repeated application of Proposition \ref{prop:desess}. Hence, by induction we have for all $1 \leq j \leq n$ that $t'_j$ mirrors $t_j$ in $P_j$, where $P_j$ is the set of position essential for $P$ in both $t'_j$ and $t_j$. Moreover, for each essential redex in $\mathcal{V}'_j$ there exists an essential redex in $\mathcal{V}_j$ at the same position and employing the same rewrite rule, and vice versa.

For all $1 \leq j \leq n$ the path prefix sets of $\mathcal{V}'_j$ and $\mathcal{V}_j$ are identical by the correspondence between the prefix sets and the essential redexes. Thus, $E/u$ mirrors $E$ in $P$ and $\pmep{P}{E/u} = \pmep{P}{E}$. By Lemma \ref{lem:simple_proj} it now follows that $D \emmy u$ mirrors $D$ in $P$ and that $\pmep{P}{D \emmy u} = \pmep{P}{D}$, as required.
\qed

With the help of the previous two lemmas and Lemma \ref{lem:simple_proj} we can extend the emaciated projection to a projection of finite sequences of complete developments across reductions of arbitrary length, taking into account, in the final terms of the finite sequences of complete developments, the residuals of the redexes projected across.

\begin{defi}
\label{def:emmybig}
Let $P$ be a prefix set of $s_n$ in $D$. If $s_0 = t_0 \trewt^\alpha t_\alpha$, then the \emph{emaciated projection} of $D$ across $t_0 \trewt^\alpha t_\alpha$ with respect to $P$, denoted $D \emmy (t_0 \trewt^\alpha t_\alpha)$, is defined as follows:
\begin{enumerate}[(1)]
\item
if $\alpha = 0$, then $D \emmy (t_0 \trewt^\alpha t_\alpha) = D$,
\item
if $\alpha = \alpha' + 1$, then  $D \emmy (t_0 \trewt^\alpha t_\alpha) = D_{\alpha'} \emmy (t_{\alpha'} \rew t_{\alpha' + 1})$ where $D_{\alpha'} = D \emmy (t_0 \trewt^{\alpha'} t_{\alpha'})$ provided $t_{\alpha'} \rew t_{\alpha' + 1}$ contracts a redex $u$ such that no redex in $u / D_{\alpha'}$ occurs at a position in $P$,
\item
if $\alpha$ is a limit ordinal, then $D \emmy (t_0 \trewt^\alpha t_\alpha)$ is defined as the result of applying Lemma \ref{lem:simple_proj} to $D_\beta = D \emmy (t_0 \trewt^\beta t_\beta)$, with $D \emmy (t_0 \trewt^\alpha t_\alpha)$ starting in $t_\alpha$ and with $\beta < \alpha$ such that for all $D_\gamma = D \emmy (t_0 \trewt^\gamma t_\gamma)$ with $\beta \leq \gamma < \alpha$ it holds that $\pmep{P}{D_\gamma} = \pmep{P}{D_\beta}$.
\end{enumerate}
\end{defi}

\noindent Hence, the emaciated projection is only defined if the condition in the successor ordinal case is satisfied for every step along $t_0 \trewt t_\alpha$. The condition in the limit ordinal case can always be satisfied. This follows by Lemmas \ref{lem:essential_then_decrease} and \ref{lem:inessential_nonroot_then_equal} and since $\prec$ is well-founded. In fact, by Lemma \ref{lem:inessential_nonroot_then_equal} and the construction in Lemma \ref{lem:simple_proj}, it follows that choosing any $D_\beta$ satisfying the desired criteria yields the same finite sequence of complete developments for $D \emmy (t_0 \trewt t_\alpha)$.

\begin{exa}
Suppose we have a fully-extended, orthogonal iCRS that has the following two rewrite rules:
\begin{align*}
f([x]Z(x)) & \rew Z(Z(a)) \\
g(Z)       & \rew h(Z)
\end{align*}
We can now define the following finite sequence of complete developments $D$, with each development consisting of a single step:
\[
g(f([x]g(g(x)))) \rew g(f([x]g(h(x)))) \rew g(f([x]h(h(x)))) \rew g(h^4(a)) \, .
\]

Consider the prefix set $P = \{ \epsilon, 1 \}$ of $g(h^4(a))$, i.e.\ the set of positions of the context $g(h(\Box))$ with exception of the position of the hole. Applying the map $\pmap$ to $D$ with respect to $P$ yields the prefix $\{ \epsilon, 1, 10, 101 \}$, i.e.\ the set of positions of the context $g(f([x]g(\Box)))$ again with exception of the position of the hole. Hence, the $f([x]Z(x)) \rew Z(Z(a))$-redex in $g(f([x]g(g(x))))$ is essential. Across that redex the emaciated projection yields the following finite sequence:
\[
g^5(a) =  g^5(a) \rewt g(h(g(h(g(a))))) = g(h(g(h(g(a))))) \, .
\]
The first complete development in the sequence becomes empty as it only contracts inessential redexes before the projection. Moreover, the last development becomes empty as the $f([x]Z(x)) \rew Z(Z(a))$-redex is a residual of the contracted redex.

If we next contract the redex at position $1111$ in $g^5(a)$, which is inessential with respect to the prefix set $P$ of $g(h(g(h(g(a)))))$, the emaciated projection yields the following finite sequence:
\[
g^4(h(a)) = g^4(h(a)) \rew g(h(g^2(h(a)))) = g(h(g^2(h(a)))) \, .
\]
The second complete development in the sequence no longer contracts the redex at position $111$, as that particular redex is inessential. 

Finally, contracting the redex at position $1$ in $g^4(h(a))$, which is essential with respect to the prefix set $P$ of $g(h(g^2(h(a))))$, the emaciated projection yields the following finite sequence:
\[
g(h(g^2(h(a)))) = g(h(g^2(h(a)))) = g(h(g^2(h(a)))) = g(h(g^2(h(a)))) \, .
\]
The second complete development in the sequence now also becomes empty as the redex at position $1$ has already been reduced.

Note that in each case the redex at the root is essential. However, the emaciated projection across contraction of this redex is undefined, as a residual of the redex occurs at a position in $P$ in the final term of each of the considered finite sequences of complete developments.
\end{exa}

We can summarise the results of this section and of the previous one in the following abstract theorem:
\begin{thm}
Let $O = \mathbb{N}^*$ be the set of finite tuples of natural numbers. Then there is
a well-founded order $\prec$ on $O$, and a pair $(\pme, \pmap)$ of maps 
such that if $D$ is a finite sequence of complete developments and $P$ is a prefix set of the final term of $D$, then: 

\begin{enumerate}[$\bullet$]
\item
$\pmep{P}{D}$ is an element of $O$, and
\item
$\pmapp{P}{D}$ is a prefix set of the initial term of $D$,
\end{enumerate}
\noindent and if $D'$ is a sequence of complete developments strictly shorter than $D$ with $P'$ a prefix set of the final term of $D'$, then $\pmep{P'}{D'} \prec \pmep{P}{D}$.

For $s \trewt t$, with $s$ the initial term of $D$, it further holds that:
\begin{enumerate}[\em(1)]
\item
if $s \trewt t$ consists of a \emph{single} step contracting a redex $u$ at a position in $\pmapp{P}{D}$, with no residual in $u/D$ occurring at a position in $P$, then there exists a $D'$ such that $\pmep{P}{D'} \prec \pmep{P}{D}$, and
\item
if $s \trewt t$ consists of one or more steps and \emph{only} contracts redexes at positions \emph{not} in $\pmapp{P}{D}$, then there exists a $D'$ such that $\pmep{P}{D'} = \pmep{P}{D}$ and $\pmapp{P}{D'} = \pmapp{P}{D}$,
\end{enumerate}
where in both cases $D'$ is a finite sequence of complete developments with initial term $t$ such that the final term of $D'$ mirrors the final one of $D$ in $P$.
\end{thm}

\proof
Let $O = \mathbb{N}^*$, that is the set of (possibly empty) tuples of natural numbers.
The well-founded order $\prec$ is obtained by comparing tuples first length-wise and next lexicographically (in the natural order), as described below Definition \ref{measuredef}. The map $\pme$ is then as in Definition~\ref{measuredef}, and $\pmap$ is the inductive extension to finite sequences of complete developments as described below Definition \ref{def:pmap}.

To see that $(\pme, \pmap)$ satisfies the two properties with respect to reductions, take the emaciated projection from Definition \ref{def:emmybig}. The first property with respect to reductions follows by Lemma \ref{lem:essential_then_decrease}; the second property follows by Lemma \ref{lem:inessential_nonroot_then_equal} in the successor ordinal case and by Lemma \ref{lem:simple_proj} in the limit ordinal case, as in Definition \ref{def:emmybig}.
\qed

Note that the two properties with respect to reductions are as claimed in Section \ref{sec:overoverview} (when extended to reductions of arbitrary length).
Moreover, 
the above theorem establishes that (sound) projection pairs, as defined and employed in \cite{K08} and \cite{paper_iv}, exist. Without giving definitions, we mention that the first part of the theorem immediately gives us existence of projection pairs, while the second part shows that the given projection pair is sound. We do not employ the device of projection pairs in the current paper as the main theorems presented below require much more fine-grained details of the maps $\pme$ and $\pmap$ than provided by projection pairs.

\subsection{Properties of essential positions and redexes}
\label{sec:ess_prop}

We prove some further properties of essential positions and essential redexes with regard to complete developments and projections. We first relate essential positions along different complete developments of the same set of redexes:
\begin{lem}
\label{lem:devsplit}
Let $s$ and $t$ be terms, $\mathcal{U}$ a set of redexes of $s$ such that $s \dev^\mathcal{U} t$, and $P$ a prefix set of $t$. If $s \dev^{\mathcal{V}_1} t' \dev^{\mathcal{V}_2} t$ with $\mathcal{V}_1 \subseteq \mathcal{U}$ and $\mathcal{V}_2 = \mathcal{U}/(s \dev t')$, then the set of positions essential for $P$ in $s$ is identical along $s \dev t$ and $s \dev t' \dev t$.
\end{lem}

\proof
By Lemma \ref{compimplfinite}, $\mathcal{U}$ satisfies the finite jumps property. Hence, since $s \dev t$ and $s \dev s' \dev t$ are both complete developments of $\mathcal{U}$, we have by Theorem \ref{fjdt} that the set of descendants in $t$ of a position in $s$ is identical along both developments. By Proposition~\ref{prop:desess} it now follows for any position in $s$ with a descendant in $P$ that the position is essential irrespective of the development being either $s \dev t$ or $s \dev s' \dev t$, where the proposition is applied twice in case of the latter development. This leaves to prove that the same holds for positions in redex patterns of redexes in $\mathcal{U}$.

Consider a fresh unary function symbol $f$ and replace each subterm $s'$ of $s$ with a redex from $\mathcal{U}$ occurring at the root by $f(s')$. This yields a term $s^f$. Since the unary function symbol $f$ does not occur in any of the rewrite rules of the assumed iCRS, it is easily shown for each (not necessarily complete) development starting in $s$ that there exists a corresponding development starting in $s^f$, where the set of redexes is adapted appropriately and such that the removal of all function symbols $f$ yields the original development. Hence, the completeness of a development starting in $s$ implies the completeness of the corresponding development starting in $s^f$.

Suppose that $s^f \dev t^f$ is the complete development that corresponds to $s \dev t$. Define the prefix set $P^f$ of $t^f$ in such a way that the removal of the function symbols $f$ from $t^f$ and the corresponding elements from the positions in $P^f$ yields $P$ and such that for any position $p \in P^f$ that is not the prefix of any another position in $P^f$ it holds that $\rs{t^f|_p} \not = f$. By definition of $P^f$ and the definition of essentiality, a redex in $\mathcal{U}$ is essential if and only if the function symbol $f$ directly preceding it in $s^f$ is. Hence, by looking at the function symbols $f$ occurring in $s^f$, the result now follows for the positions in the redex patterns of the redexes in $\mathcal{U}$ in similar fashion as for all other positions.
\qed

By the previous lemma we immediately have the following:
\begin{cor}
\label{cor:devsplit}
Let $s$ and $t$ be terms, $\mathcal{U}$ a set of redexes of $s$ such that $s \dev^\mathcal{U} t$, and $P$ a prefix set of $t$. If it holds that:
\begin{enumerate}[$\bullet$]
\item
$s \dev^{\mathcal{V}_1} s' \dev^{\mathcal{V}_2} t$ with $\mathcal{V}_1 \subseteq \mathcal{U}$ and $\mathcal{V}_2 = \mathcal{U}/(s \dev s')$, and
\item
$s \dev^{\mathcal{V}'_1} t' \dev^{\mathcal{V}'_2} t$ with $\mathcal{V}'_1 \subseteq \mathcal{U}$ and $\mathcal{V}'_2 = \mathcal{U}/(s \dev t')$,
\end{enumerate}
then the set of positions essential for $P$ in $s$ is identical along $s \dev s' \dev t$ and $s \dev t' \dev t$.
\end{cor}

We next show that each essential redex has an essential residual as long as it is not contracted and that inessential redexes only have inessential residuals. Moreover, we show that the same holds in case emaciated projections are considered.

\begin{lem}
\label{lem:essstaysess_prim}
Let $D: s_0 \dev s_1 \dev \cdots \dev s_n$ and let $P$ be a prefix set of $s_n$. If $s_0 \rew t_0$ contracts a redex $u$ such that no redex in $u/D$ occurs at a position in $P$ and such that $D/u$ exists, then for every redex $v$ in $s_0$:
\begin{enumerate}[$\bullet$]
\item
if $v$ is essential, then $v$ is either $u$ or there exists a residual of $v$ in $t_0$ that is essential for $P$ along $D/u$, and
\item
if $v$ is inessential, then all residuals of $v$ in $t_0$ are inessential for $P$ along $D/u$.
\end{enumerate}
\end{lem}

\proof
Consider the following diagram:
\[
\xymatrix{
s_0 \ar@{=>}[r] \ar[d]^{u}
  & s_1 \ar@{=>}[r] \ar@{=>}[d]
  & \cdot \ar@{.}[r] \ar@{=>}[d]
  & \cdot \ar@{=>}[r]
  & s_n \ar@{=>}[d]^{u/D} \\
t_0 \ar@{=>}[r]
  & t_1 \ar@{=>}[r]
  & \cdot \ar@{.}[r]
  & \cdot \ar@{=>}[r]
  & t_n
}
\]
where the reduction at the bottom is $D/u$. Since no redex in $u/D$ occurs at a position in $P$, we have that $t_n$ mirrors $s_n$ in $P$. Hence, we can consider the redexes in $t_0$ that are essential for $P$. By repeated application of Corollary \ref{cor:devsplit} to the tiles of the diagram, it follows for all $0 \leq i < n$ that the set of essential positions in $s_i$ is the identical along both $s_i \dev s_{i + 1}$ and $s_i \dev t_i \dev t_{i + 1}$ when we consider the positions essential for $P$ in $s_{i + 1}$ along $s_{i + 1} \dev^* s_n$ and in $t_{i + 1}$ along $t_{i + 1} \dev^* t_n$, respectively. Hence, the result follows by Proposition \ref{prop:desess}.
\qed

\begin{lem}
\label{lem:essstaysess}
Let $D : s_0 \dev s_1 \dev \cdots \dev s_n$ and let $P$ be a prefix set of $s_n$. If $s_0 \rew t_0$ contracts a redex $u$ such that no redex in $u/D$ occurs at a position in $P$, then for every redex $v$ in $s_0$:
\begin{enumerate}[$\bullet$]
\item

if $v$ is essential, then $v$ is either $u$ or there exists a residual of $v$ in $t_0$ that is essential for $P$ along $D \emmy u$, and
\item
if $v$ is inessential, then all residuals of $v$ in $t_0$ are inessential for $P$ along $D \emmy u$.
\end{enumerate}
\end{lem}

\proof
By Lemma \ref{lem:simple_proj} a redex is essential for $P$ along the finite sequence of complete developments $E$ obtained through the lemma when starting in $s_0$ if and only if the is essential for $P$ along $D$. Hence, the result follows by application of the Lemma \ref{lem:essstaysess_prim} to $E$.
\qed

\subsection{Reductions to normal form}
\label{sec:nf_emmy}

We show that the result of an emaciated projection is always defined in case the finite sequence of complete developments that is projected is in fact the beginning of a reduction to normal form, i.e.\ the beginning of a reduction to a term without redexes. It is important to note that any finite reduction can be seen as a finite sequence of complete developments: Simply assume that each development consists of a single step.

To show the result, we first establish that the emaciated projections of finite sequences of complete developments mirror each other in some prefix set in case the finite sequences themselves mirror each other in that prefix set. Moreover, we establish a relation between the emaciated projections of a finite sequence of complete developments with respect to different prefix sets.

\begin{lem}
\label{lem:nf_mirror}
Let $D$ and $E$ be finite sequences of complete developments starting in the same term $s$. Moreover, let $P$ be a prefix set of the final term of $D$ and let $E$ mirror $D$ in $P$. If $s \rew t$ contracts a redex $u$ such that no redex in either $u / D$ or $u / E$ occurs at a position in $P$, then $E \emmy u$ mirrors $D \emmy u$ in $P$.
\end{lem}

\proof
Suppose $s \rew t$ contracts a redex $u$ such that no redex in either $u / D$ or $u / E$ occurs at a position in $P$. Let $D'$ and $E'$ be the result of applying Lemma \ref{lem:simple_proj} to $D$ and $E$, respectively, with $D'$ and $E'$ both starting in $s$. By construction and since $E$ mirrors $D$ in $P$, it follows that $D'$ and $E'$ are identical. Hence, $D' / u$ and $E' / u$ are identical and, by the definition of the emaciated projection, $E \emmy u$ mirrors $D \emmy u$ in $P$.
\qed

To establish a relation between the emaciated projection of a finite sequence of complete development with respect to different prefix sets we define an extension of mirroring. In the definition we write $D$, respectively $E$, for the finite sequence of complete developments $\fscd{s}{U}{n}$, respectively $\fscd{t}{V}{n}$, of length $n$. Moreover, for all $0 \leq i \leq n$ we denote by $P_i$, respectively $Q_i$, the set of positions essential for a prefix set $P$ in $s_i$, respectively for a prefix set $Q$ in $t_i$.

\begin{defi}
Let $P$ and $Q$ with $Q \subseteq P$ be prefix sets of, respectively, $s_n$ in $D$ and $t_n$ in $E$. The finite sequence $E$ \emph{sub-mirrors} $D$ in $Q \subseteq P$ if for all $0 \leq i \leq n$ it holds that $Q_i \subseteq P_i$, $t_i$ mirrors $s_i$ in $Q_i$, and the path prefix set of $Q_i$ with respect to $\mathcal{V}_i$ is a subset of the path prefix set of $P_i$ with respect to $\mathcal{U}_i$.
\end{defi}

We can now relate the emaciated projection of a finite sequence of complete developments with respect to different prefix sets in case sub-mirroring holds.
\begin{lem}
\label{lem:nf_submirror}
Let $D$ be a finite sequence of complete developments starting in a term $s$. Moreover, let $P$ and $P'$ be prefix sets of the final term of $D$ such that $P' \subseteq P$. If $s \rew t$ contracts a redex $u$ such that no redex in $u / D$ occurs at a position in $P$, then $D \emmy u$ with respect to $P'$ sub-mirrors $D \emmy u$ with respect to $P$.
\end{lem}

\proof
Suppose $s \rew t$ contracts a redex $u$ such that no redex in $u / D$ occurs at a position in $P$. Hence, since $P' \subseteq P$, no redex in $u / D$ occurs at a position in $P'$ either. Let $E$ and $E'$ be the result of applying Lemma \ref{lem:simple_proj} to $D$ with respect to $P$ and $P'$, respectively, such that $E$ and $E'$ start in $s$. By construction, $E'$ sub-mirrors $E$ in $P' \subseteq P$. Moreover, no redex in $u /E$, respectively $u / E'$, occurs at a position in $P$, respectively in $P'$.

Let $s_0 = s'_0 = s$ and $t_0 = t'_0 = t$ and write $E : \fscd{s}{U}{n}$ and $E' : \fscdp{s}{U}{n}$. Consider the following two diagrams, which are the result of iteratively applying Proposition \ref{prop:finite_over_normal} to $E$ and $E'$, respectively.
\[
\xymatrix@=0.78cm{
s_0 \ar@{=>}[r] \ar[d]^{u}
  & s_1 \ar@{=>}[r] \ar@{=>}[d]
  & \cdot \ar@{.}[r] \ar@{=>}[d]
  & \cdot \ar@{=>}[r]
  & s_n \ar@{=>}[d]^{u / E}
&
s'_0 \ar@{=>}[r] \ar[d]^{u}
  & s'_1 \ar@{=>}[r] \ar@{=>}[d]
  & \cdot \ar@{.}[r] \ar@{=>}[d]
  & \cdot \ar@{=>}[r]
  & s'_n \ar@{=>}[d]^{u / E'}
 \\
t_0 \ar@{=>}[r]
  & t_1 \ar@{=>}[r]
  & \cdot \ar@{.}[r]
  & \cdot \ar@{=>}[r]
  & t_n
&
t'_0 \ar@{=>}[r]
  & t'_1 \ar@{=>}[r]
  & \cdot \ar@{.}[r]
  & \cdot \ar@{=>}[r]
  & t'_n
}
\]
Let $P_i$, respectively $P'_i$, be the set of positions essential for $P$ in $s_i$, respectively for $P'$ in $s'_i$. Since $E'$ sub-mirrors $E$ in $P' \subseteq P$, it follows for all $0 \leq i \leq n$ that $P'_i \subseteq P_i$. Moreover, for each redex in $u / (s'_0 \dev^* s'_i)$ that occurs at a position in $P'_i$ it follows that there exists a redex in $u / (s_0 \dev^* s_i)$ that occurs at the same position. In addition, it holds for each such redex that all positions in its redex pattern occur in $P'_i$, otherwise a redex in $u / E'$ occurs at a position in $P'$.

By induction, employing the definition of descendants and the above facts, it follows that $t'_i$ mirrors $t_i$ in $P'_i / (s'_i \dev t'_i) \subseteq P_i / (s_i \dev t_i)$. Moreover, by fully-extendedness and orthogonality we have for each redex in $\mathcal{U}'_i / (s'_i \dev t'_i)$ that there exists a redex in $t_i$ at the same position and employing the same rewrite rule. And, since $E'$ sub-mirrors $E$ in $P' \subseteq P$, it also follows for each redex in $\mathcal{U}_i / (s_i \dev t_i)$ that if all positions in its redex pattern occur in $P_i / (s_i \dev t_i)$ but some do not occur in $P'_i / (s'_i \dev t'_i)$, then in fact no position of the redex pattern occurs in  $P'_i / (s'_i \dev t'_i)$.

Since $t_n$ mirrors $s_n$ in $P$, respectively $t'_n$ mirrors $s'_n$ in $P'$, it follows by induction, applying Proposition \ref{prop:desess}, that $P_i / (s_i \dev t_i)$, respectively $P'_i / (s'_i \dev t'_i)$, is in fact the set of position essential for $P$ in $t_i$, respectively the set of position essential for $P'$ in $t'_i$. Hence, $E' / u$ sub-mirrors $E / u$ in $P' \subseteq P$. That $D \emmy u$ with respect to $P'$ sub-mirrors $D \emmy u$ with respect to $P$ now follows by definition of the emaciated projection.
\qed

Finally, we can prove the main result of this section.
\begin{lem}
\label{lem:nf_ok}
Let $s_0 \rewt s_1 \rewt \cdots \rewt s_d \rewt \cdots \, s_\omega$ be a reduction to normal form of length at most $\omega$ such that, for all $d \in \natnum$, the steps in $s_d \trewt s_\omega$ occur below depth $d$ and $P_d$ is the set of positions in $s_d$ above depth $d$. For every $s_0 \trewt t_0$, redex $u$ in $t_0$, and $d \in \natnum$ no redex in $u / D_d$ occurs at a position in $P_d$, with $D_d = (s_0 \rewt s_d) \emmy (s_0 \trewt t_0)$ with respect to~$P_d$.
\end{lem}

\proof
We reason by contradiction. Thus, suppose a reduction $s_0 \trewt t_0$, redex $u$ in $t_0$, and $d \in \natnum$ exist such that a redex in $u / D_d$ occurs at a position in $P_d$, with $D_d = (s \rewt s_d) \emmy (s_0 \trewt t_0)$. Note that $D_d$ is defined, otherwise there is a reduction shorter than $s_0 \trewt t_0$ satisfying the required conditions.

Assume $s \rewt s_d$ has length $n$. Moreover, let $k$ be the maximum value $|p|$ with $p$ a position in the redex pattern of the rule employed in $u$ and assume $s \rewt s_{d + k}$ has length~$m$.

Write $E_d$, respectively $E_{d + k}$, for $(s_0 \rewt s_{d + k}) \emmy (s_0 \trewt t_0)$ with respect to $P_d$, respectively $P_{d + k}$. By ordinal induction on the length of $s \trewt t$, applying Lemma \ref{lem:nf_mirror} with respect to $P_d$ and Lemma \ref{lem:nf_submirror} with respect to $P_d \subseteq P_{d + k}$, it follows that $E_d$ sub-mirrors $E_{d + k}$ in $P_d \subseteq P_{d + k}$. Moreover, by definition of emaciated projections and since all steps in $s_d \rewt s_{d + k}$ occur below depth $d$, we have that $D_d$ mirrors the sequence of the first $n$ developments of $E_d$ in $P_d$ and that all steps in the last $m - n$ developments of $E_d$ and $E_{d + k}$ occur below depth $d$. By orthogonality and fully-extendedness, it follows for the redex in $u / D_d$, which occurs at a position $p \in P_d$ in the final term of $D_d$, that a redex in $u / E_d$ occurs at $p$ in the final term of $E_d$. Moreover, since $E_d$ sub-mirrors $E_{d + k}$, this implies that a redex in $u / E_{d + k}$ occurs at $p$ in the final term of $E_{d + k}$. But then, since the final term of $E_{d + k}$ mirrors $s_{d + k}$ in $P_{d + k}$, it holds for $s_{d + k}$ that a redex and all the positions in its redex pattern occur in $P_{d + k}$, which is impossible by definition of $s_{d + k}$, contradiction.
\qed

\section{Fair reduction strategies}
\label{sec:normal}

In this section we consider reduction strategies in fully-extended, orthogonal iCRSs. We show that all considered reduction strategies are normalising, where a normal form is understood as usual:
\begin{defi}
A term in an iCRS is a \emph{normal form} if no redexes occur in the term.
\end{defi}

We consider outermost-fair, fair, and needed-fair reduction strategies. These strategies are `fair' in the sense that each assumes a special class of redexes $\mathcal{P}$ and ensures for every redex in $\mathcal{P}$ that if one occurs in a term along a reduction, then after a finite number of further steps in the reduction, either a residual of the redex is contracted, or no residuals of the redex are in $\mathcal{P}$ . Formally, the reduction strategies satisfy the following definition.
\begin{defi}
\label{def:pfair}
Let $\mathcal{P}$ be a predicate. A \emph{$\mathcal{P}$-fair reduction} is a weakly continuous reduction $(s_\beta)_{\beta < \alpha}$ where for every $\beta < \alpha$ and redex $u$ in $s_\beta$ satisfying $\mathcal{P}$ there exists a $\beta \leq \gamma < \min ( \alpha, \beta + \omega )$ such that either
\begin{enumerate}[(1)]
\item
$s_\gamma \rew s_{\gamma + 1}$ contracts a residual of $u$ satisfying $\mathcal{P}$, or
\item
no residual of $u$ in $s_\gamma$ satisfies $\mathcal{P}$.
\end{enumerate}
\end{defi}

Thus, each redex that satisfies $\mathcal{P}$ is either reduced after a finite number of steps or it no longer satisfies $\mathcal{P}$ after a finite number of steps.

\begin{rem}
The condition that a fair reduction should be weakly continuous is simply a safeguard against reduction strategies that are normalising only by cheating. For example, in the iCRS which has the rules $a \rew a$ and $b \rew f(b)$, the reduction
\[
a \rew a \rew \cdots \, b \rew f(b) \rew \cdots \, f^\omega
\]
is normalising, but none of the terms in the first $\omega$ steps bear any relationship to the final term as the reduction is not weakly continuous, let alone convergent.
\end{rem}

\begin{exa}
\label{ex:outfair}
Given that a redex at a position $p$ is called outermost if no redex occurs at a strict prefix position of $p$, we can define \emph{outermost-fair} reductions by defining a predicate on redexes that is true in case the redex is outermost and false otherwise. Consider the following two rewrite rules:
\begin{align*}
f(Z) & \rew g(Z) \\
a    & \rew g(a)
\end{align*}
Next, consider the reductions
\[
f(a) \rew f(g(a)) \rew \cdots \rew f(g^n(a)) \rew g^{n + 1}(a) \rew \cdots \, g^\omega
\]
and
\[
f(a) \rew f(g(a)) \rew \cdots \rew f(g^n(a)) \rew \cdots \, f(g^\omega) \rew g^\omega \, .
\]
The first of these reductions is outermost-fair, as a residual of each redex present in each term is reduced after a finite number of steps. The second reduction is \emph{not} outermost-fair, as a residual of a redex that occurs at the root of $f(a)$ is only contracted after $\omega$ steps and as a redex occurring at the root of a term is always outermost.
\end{exa}

\subsection{Outermost-fair reductions}
\label{sec:outfair}

The standard way of obtaining normal forms in (finitary)
higher-order rewriting is by using an \emph{outermost-fair} strategy
\cite{OD77,R97,O99}. 

\begin{defi}
Let $s$ be a term. A redex at a position $p$ in $s$ is \emph{outermost} if no redex in $s$ occurs at a strict prefix position of $p$. 
The predicate $\Poutermost$ is satisfied by a redex if{f} that redex is outermost.
 An \emph{outermost-fair reduction} is a $\Poutermost$-fair reduction.
\end{defi}

Hence, a reduction is outermost-fair if, after a finite number of steps, every outermost redex is either reduced or not outermost anymore.

Outermost-fair reductions satisfy the following property, the proof of which is similar to the one of Theorem 3 in \cite{O99}.
\begin{lem}
\label{lem:outermost_help}
Let $s$ be a term and $T$ an outermost-fair reduction of length at least $\omega$ starting in $s$. If there is a reduction $s \trewt t$ to normal form, then $T$ is strongly convergent of length $\omega$ with $t$ as its final term.
\end{lem}

\proof
By compression, we may assume that $s \trewt t$ has length at most $\omega$. Moreover, by strong convergence we may write $s \trewt t$ as:
\[
s \rewt s_1 \rewt \cdots \rewt s_d \rewt \cdots \, t \, ,
\]
where all steps in $s_d \trewt t$ occur below depth $d$. For each depth $d > 0$ and $D_d : s \rewt s_d$, we have by definition that $s_d$ mirrors $t$ in $P_d$, where $P_d$ is the set of positions in $t$ above depth $d$. Moreover, no redexes occur in $s_d$ at positions in $P_d$ and, as $s \rewt s_d$ is finite, we can view $D_d$ to be a finite sequence of complete developments, where each development consists of a single step.

Let the depth $d > 0$ be arbitrary and denote the first $\omega$ steps of $T$ by $T_\omega$. Consider $D_d \emmy T_\omega$, which exists by Lemma \ref{lem:nf_ok}. By Lemmas~\ref{lem:essential_then_decrease} and \ref{lem:inessential_nonroot_then_equal} and well-foundedness of $\prec$, only a finite number of steps of $T_\omega$ is essential for $P_d$. Following the finite number of essential steps, there are two possibilities for the emaciated projection of $D_d$ by the construction in Lemma \ref{lem:simple_proj}: Either all developments in the projection are empty, or not.
\begin{enumerate}[$\bullet$]
\item
In case all developments are empty it follows by Lemma \ref{lem:inessential_nonroot_then_equal} that all remaining terms along $T_\omega$ mirror $s_d$ and $t$ in $P_d$ and that no redexes are contracted above depth $d$.
\item
 In case not all developments are empty, it follows by Lemma \ref{lem:inessential_nonroot_then_equal} that there exists a fixed set of essential positions $P$ such that all the remaining terms along $T_\omega$ mirror each other in $P$. Moreover, the lemma together with non-emptiness implies that a redex $u$ occurs at a fixed position in $P$. Since the depth of $u$ is finite, only a finite number of redexes can be created above $u$ in the remaining part of $T_\omega$. These redexes cannot be contracted or cease to exist by orthogonality and since all further contracted redexes occur at positions not in $P$, again by Lemma \ref{lem:inessential_nonroot_then_equal}. Hence, after a finite number of further steps a redex must be created that is outermost for the remainder of $T_\omega$, contradicting outermost-fairness. Thus, the emaciated projection of $D_d$ must become empty after a finite number of steps.
\end{enumerate}

As the previous holds for all depths $d > 0$, we have that $T_\omega$ is strongly convergent with limit $t$. Hence, $T = T_\omega$ and the result follows.
\qed

We thus obtain a strong result concerning normalisation of iCRSs:

\begin{thm}
\label{thm:outermostfair}
If $s$ can be reduced to normal form by a strongly convergent
reduction, then it also reduces to a normal form by
any outermost-fair reduction. Any
such reduction is strongly convergent and of length at most $\omega$.
\end{thm}

\proof
If $T$ is a finite outermost-fair reduction starting in $s$
and $T$ reaches a normal form, then we are done. If $T$ is finite
but has not reached a normal form, then there is at least one
outermost redex in the final term of $T$, and we may thus extend it.
Hence, we only need to prove that if $T$ is infinite, then $T$ is
strongly convergent of length $\omega$ and reaches a normal form.
This is the content of Lemma \ref{lem:outermost_help}.
\qed

\subsection{Fair reductions}
\label{sec:fair}

Contrary to the predicate considered in the previous section, which is only satisfied under certain conditions, this section considers a predicate that is always satisfied.
\begin{defi}
Let $\Ptrue$ be the predicate that is true on all redexes in all terms.
A \emph{fair reduction} is a $\Ptrue$-fair reduction.
\end{defi}

As the predicate is always true, the second clause of Definition \ref{def:pfair} cannot occur in a fair reduction unless the first clause applies earlier in the considered reduction.

We have the following:
\begin{thm}
\label{thm:fair}
If $s$ can be reduced to a normal form by a strongly convergent reduction, then it also reduces to a normal form by any fair reduction. Any such reduction is strongly convergent and of length at most $\omega$.
\end{thm}

\proof
Since a fair reduction is in particular fair with respect to outermost redexes, the result follows by Theorem \ref{thm:outermostfair}. 
\qed

As each fair reduction is an outermost-fair reduction, it follows that the predicate that is true on all redexes in all terms strengthens the predicate used for outermost-fair reductions. Hence, we obtained a weaker result than in the previous section.

\subsection{Needed-fair reductions}
\label{sec:needfair}

In the vein of Huet and L\'evy \cite{HL91}, who show that needed reductions are normalising for term rewriting systems, we next prove that needed-fair reductions are normalising for iCRSs.

\begin{defi}
Let $s$ be a term. A redex $u$ in $s$ is \emph{needed} if, along every strongly convergent reduction from $s$ to a normal form, some residual of $u$ is contracted. The predicate $\Pneeded$ is satisfied by a redex if{f} that redex is needed. A \emph{needed-fair reduction} is a $\Pneeded$-fair reduction.
\end{defi}

By definition of neededness, the second clause of Definition \ref{def:pfair} cannot occur in a needed-fair reduction unless the first clause applies earlier on in the considered reduction, otherwise the considered redex is not needed.

\begin{exa}
Consider the rewrite rules and reductions from Example \ref{ex:outfair}. The first reduction is needed-fair, as each redex along the reduction is reduced after a finite number of steps. The second reduction is not needed-fair, as the redex at the root of $f(a)$ is only reduced after $\omega$ steps and as the redex at the root of a term is by definition needed.
\end{exa}

To prove normalisation of needed-fair reductions, we establish a relation between essential redexes and needed redexes.
\begin{lem}
Let $s_0 \rewt s_1 \rewt \cdots \rewt s_d \rewt \cdots \, s_\omega$ be a reduction to normal form of length at most $\omega$ such that, for all $d \in \natnum$, the steps in $s_d \trewt s_\omega$ occur below depth $d$ and $P_d$ is the set of positions in $s_d$ above depth $d$. If a redex in $s_0$ is essential for some $P_d$ in $s_0$ with $d \in \natnum$, then the redex is needed.
\end{lem}

\proof
We reason by contradiction. Thus, suppose $u$ is a redex in $s_0$ that is essential for some prefix set $P_d$ but not needed. By definition of neededness there exists a reduction $s_0 = t_0 \trewt t_\alpha$ to normal form that does not contract any residual of $u$. Write $D_\alpha = (s_0 \rewt s_d) \emmy (t_0 \trewt t_\alpha)$, which exists by Lemma \ref{lem:nf_ok} and since $s_0 \rewt s_d$ can be seen as a finite sequence of complete developments, each consisting of a single step. We show by ordinal induction that a residual of $u$ occurs in $t_\alpha$, the first term of $D_\alpha$, that is essential for $P_d$. Obviously, for $t_0 = s_0$ the result is immediate by assumption.

For $t_{\alpha + 1}$ the result follows by the induction hypothesis and Lemma \ref{lem:essstaysess}, since by assumption the redex contracted in $t_\alpha \rew t_{\alpha + 1}$ is not a residual of $u$.

For $t_\alpha$, with $\alpha$ a limit ordinal, the result follows by the induction hypothesis and strong convergence, since for some $\beta < \alpha$ we have by definition of the emaciated projection that $D_\alpha$ mirrors $(s_0 \rewt s_d) \emmy (t_0 \trewt t_\beta)$ in $P_d$.

Hence, a residual of $u$ occurs in $t_\alpha$, contradicting that $t_\alpha$ is a normal form. Hence, every redex that is essential for some $P_d$ is needed.
\qed

\begin{lem}
Let $s_0 \rewt s_1 \rewt \cdots \rewt s_d \rewt \cdots \, s_\omega$ be a reduction to normal form of length at most $\omega$ such that, for all $d \in \natnum$, the steps in $s_d \trewt s_\omega$ occur below depth $d$ and $P_d$ is the set of positions in $s_d$ above depth $d$. If a redex in $s_0$ is needed, then there exists a $d \in \natnum$ such that the redex is essential for $P_d$ in $s_0$.
\end{lem}

\proof
Consider a reduction $S$ that contracts for increasingly larger $d \in \natnum$ all redexes essential for $P_d$, considering the emaciated projection of $s_0 \rewt s_d$, until such redexes no longer occur. It is possible to employ the emaciated projection by Lemma \ref{lem:nf_ok}. By Lemmas~\ref{lem:essential_then_decrease} and \ref{lem:inessential_nonroot_then_equal} and the well-foundedness of $\prec$ it follows that $S$ is of length at most $\omega$. Moreover, again by Lemma \ref{lem:nf_ok} --- considering the emaciated projections of $s_0 \rewt s_e$ for all $e$ smaller than the $d$ under consideration at some point --- it follows that $S$ is strongly convergent. Hence, by Lemma \ref{lem:essstaysess} a redex can only be needed in $s_0$ if there exists some $d \in \natnum$ such that the redex is essential for $P_d$ in $s_0$.
\qed

Combining the above two lemmas, we obtain:
\begin{lem}
\label{lem:essisneed}
Let $s_0 \rewt s_1 \rewt \cdots \rewt s_d \rewt \cdots \, s_\omega$ be a reduction to normal form of length at most $\omega$ such that, for all $d \in \natnum$, the steps in $s_d \trewt t$ occur below depth $d$ and $P_d$ is the set of positions in $s_d$ above depth $d$. A redex in $s_0$ is needed if{f} there exists a $d \in \natnum$ such that the redex is essential for $P_d$ in $s_0$.
\end{lem}

Finally, we can prove the main theorem of this section:
\begin{thm}
\label{thm:neededfair}
If $s$ can be reduced to a normal form by a strongly convergent reduction, then it also reduces to a normal form by any needed-fair reduction. Any such reduction is strongly convergent and of length at most $\omega$.
\end{thm}

\proof
By compression, we may assume we have a reduction $s \trewt t$ to normal form of length at most length $\omega$. Write the reduction as
\[
s = s_0 \rewt s_1 \rewt \cdots \rewt s_d \rewt \cdots \, t
\]
with all steps in $s_d \trewt t$ occurring below depth $d$. Denote by $P_d$ the set of positions in $s_d$ above depth $d$.

Consider the emaciated projection of a reduction $s \rewt s_d$ with respect to a prefix set $P_d$, which is possible by Lemma \ref{lem:nf_ok}. It follows by Lemma \ref{lem:essisneed} and the needed-fair condition that an essential redex is contracted after a finite number of steps as long as any essential redexes are left. As $\prec$ is well-founded, we now have by Lemmas \ref{lem:essential_then_decrease} and \ref{lem:inessential_nonroot_then_equal} that the needed-fair reduction reduces $s$ to a term that mirrors $s_d$ in $P_d$ in a finite number of steps. Since the previous holds for any $s_d$ and $P_d$, it follows that any needed-fair reduction is a strongly convergent reduction to normal form of length at most $\omega$.
\qed

\subsection{Examples}

We proceed to show examples of fair, outermost-fair and needed-fair reductions in iTRSs and \iLC.

\begin{example}\label{ex:normal_one}
Consider the (first-order) iCRS consisting of the following three rules:
\begin{align*}
f(X,Y) & \rew g(X, f(X, Y)) \\
a      & \rew b \\
c      & \rew c
\end{align*}
and consider the term $f(a,c)$. The reduction
\[
f(a, c) \rew g(a ,f(a,c)) \rew g(b,f(a,c)) \rew g(b,g(a,f(a,c))) \rew \cdots \, ,
\]
which alternates between contracting an $f(X, Y) \rew g(X, f(X, Y))$-redex and an $a \rew b$-redex, is easily shown to be outermost-fair and needed-fair contracting only outermost redexes. The reduction is not fair as no descendant of a $c \rew c$-redex is ever reduced.

Now consider the reduction
\[
f(a, c) \rew f(b, c) \rew g(b ,f(b,c)) \rew g(b,g(b,f(b,c))) \rew g(b, g(b, g(b, f(b, c)))) \rew \cdots \, ,
\]
which first contracts the $a \rew b$-redex and then keeps contracting $f(X, Y) \rew g(X, f(X, Y))$-redexes; this reduction is easily shown to be outermost-fair and needed-fair, contracting only needed redexes. Note that the first redex contracted is not outermost and that the reduction is not fair as no descendant of a $c \rew c$-redex is ever contracted.

Finally, consider the reduction
\[
f(a, c) \rew f(b, c) \rew f(b, c) \rew g(b,f(b,c)) \rew g(b,f(b,c)) \rew g(b,g(b,f(b,c))) \rew \cdots \, ,
\]
which first contracts the $a \rew b$-redex and then alternates between contracting a $c \rew c$-redex and an $f(X, Y) \rew g(X, f(X, Y))$-redex; this reduction is easily shown to be fair, outermost-fair, and needed-fair.

All three reductions are strongly convergent, converging to the infinite term given by $s = g(b, s)$, which is clearly a normal form.
\end{example}

Recall from \cite{KOR93,T03} that $\lambda$-calculus with $\beta$-reduction 
can be fully and faithfully modelled as a CRS; likewise, \iLC with $\beta$-reduction can be modelled fully and faithfully by an iCRS \cite{JJ05a,paper_i}. The standard way of doing so is by introducing explicit symbols $\mathtt{app}$ and $\mathtt{abs}$ for application and abstraction, respectively, and defining the $\beta$-rule as
\[
\mathtt{app}(\mathtt{abs}([x]Z(x)),Z') \rightarrow Z(Z')
\]

By the above encoding we obtain the first normalising reduction strategies for \iLC. Note that we have a slight shift compared to any reduction strategy from (finite) $\lambda$-calculus: The strategies presented above will not only reduce terms with finite normal forms to their respective normal forms; they will do the same for terms that have infinite normal forms (reachable by strongly convergent reductions) even if these do not have finite normal forms.

\newlength{\omegawidth}
\settowidth{\omegawidth}{$\Omega$}

\begin{example}
Let $b$ and $g$ be variables and define the following shorthands:
\begin{align*}
h      & = \lambda w . \lambda x . \lambda y . g \, x \, (w \, x \, y) &
c      & = \Omega \hskip-\omegawidth \phantom{Y \, h} = (\lambda x. x \, x) (\lambda x . x \, x) \\
a      & = (\lambda x . x) \, b &
f      & = Y \, h = (\lambda z . (\lambda x. z \, (x \, x)) \, (\lambda x . z \, (x \, x))) \, h
\end{align*}
Consider the term $f \, a \, c$. We now recast the third reduction from Example \ref{ex:normal_one} in the iCRS representing \iLC by the standard encoding above. We employ the syntax of $\lambda$-calculus for clarity.
\begin{align*}
f \, a \, c & \rew^{\phantom{*}} f \, b \, c \\
& \rew^{\phantom{*}} f \, b \, c \\
& \rewt g \, b \,  (((\lambda x. h \, (x \, x)) \lambda x . h \, (x \, x)) \, b \, c) \\
& \rew^{\phantom{*}} g \, b \, (((\lambda x. h \, (x \, x)) \lambda x . h \, (x \, x)) \, b \, c) \\
& \rewt g \, b \, (g \, b \, (((\lambda x. h \, (x \, x)) \lambda x. h \, (x \, x)) \, a \, c)) \\
& \rew^{\phantom{*}} \cdots
\end{align*}
The reduction, where in every other step the redex in $c$ is contracted, is fair, outermost-fair, and needed-fair.
\end{example}

\section{Conclusion and suggestions for future work}
\label{sec:conclusion}

We have shown that several well-known reduction strategies are normalising for fully-extended, orthogonal infinitary Combinatory Reduction Systems (iCRSs). The proofs crucially employ the method of essential rewrite steps adapted to the infinitary setting; this method has proven to be of further use in \cite{paper_iv} where we show confluence modulo the identification of hypercollapsing subterms for fully-extended, orthogonal iCRSs.

Our results subsume identical results from first-order infinitary rewriting \cite{KKSV95}, as any iTRS can be seen as an iCRS. Moreover, we provide the first normalising reduction strategies for infinitary $\lambda$-calculus (\iLC), as \iLC can be seen as a particular example of an iCRS.

While we have provided proof of certain strategies being normalising, a full-fledged account of standardisation and the allied notion of reduction equivalence \cite{T03_OV_equiv} is still lacking and should be provided by future research. In addition, we conjecture that most of our results carry over to fully-extended, \emph{weakly} orthogonal iCRSs.

\section*{Acknowledgement}

The authors extend their thanks to the anonymous referees for their diligent work and many comments
that have led to substantial improvements in the readability of the paper.

\bibliographystyle{abbrv}
\bibliography{icrs}

\newpage

\appendix

\section{Proof of Proposition \ref{prop:phibiject}}\label{app:proof_III}

We prove Proposition \ref{prop:phibiject}.

\proof
As each path projection derives from a path, we have by definition
that $\phi$ is \emph{surjective}. Similar for the path projections
in $\mathcal{P}(s, \mathcal{U})$ and the maximal paths, as each path projection
in $\mathcal{P}(s, \mathcal{U})$ derives from a maximal path.

To prove that $\phi$ is \emph{injective}, suppose there exist (maximal) paths
$\apath, \apath'$ such that $\phi(\apath) = \phi(\apath')$. By definition
of $\phi$ both paths and the path projection consist of the same number of
nodes and edges. Let $\apath^*$ be the longest shared prefix of $\apath$ and
$\apath'$. The prefix $\apath^*$ is non-empty, as any path of $s$ starts
with $(s, \epsilon)$. There are now two cases to consider depending on
$\apath^*$ ending in either an edge or a node.

In case $\apath^*$ ends in an edge, the next node is uniquely determined
by the definition of paths. Hence, as $\apath$ and $\apath'$ have the
same number of nodes and edges we can extend $\apath^*$ with that unique node,
contradiction.

In case $\apath^*$ ends in a node, both paths extend $\apath^*$,
otherwise $\apath = \apath'$ or the paths differ in the number of nodes
or edges. In case the extension is with an unlabelled edge in case of
one of the paths, the other path must also extend $\apath^*$ with an
unlabelled edge. This follows by the definition of paths. In case
the extension is with an edge labelled $i$, the other path must also
extend $\apath^*$ with an edge labelled $i$. This follows by definition
of paths and by $\phi(\apath) = \phi(\apath')$. Hence, in case
$\apath^*$ ends in a node a contradiction also follows. We can
conclude that $\phi$ is an injection both between paths and path projections
and between maximal paths and the path projections in
$\mathcal{P}(s, \mathcal{U})$.
\qed

\end{document}